\documentclass[12pt]{article}

\usepackage{newtxtext,newtxmath}

\usepackage{graphicx}

\usepackage[letterpaper,margin=1in]{geometry}

\renewenvironment{abstract}
	{\quotation}
	{\endquotation}

\date{}

\makeatletter
\renewcommand{\fnum@figure}{\textbf{Figure \thefigure}}
\renewcommand{\fnum@table}{\textbf{Table \thetable}}
\makeatother

\usepackage{scicite}

\usepackage{url}

\def\scititle{
Quantum Co-Magnetometer Using Diamond Nitrogen-Vacancy Centers and Rubidium Cells
}
\title{\bfseries \boldmath \scititle}

\author{
	I.~Shalev$^{1\ast}$,
	K.~Levi$^{1}$,
    R.~Malkinson$^{1}$,
    A.~Hen$^{1}$,
    L.~Stern$^{1,2}$,
    N.~Bar-Gill$^{1,2,3}$\and
	\small$^{1}$Inst. of Applied Physics, The Hebrew University of Jerusalem, Jerusalem 91904, Israel.\and
	\small$^{2}$The Center for Nanoscience and Nanotechnology, The Hebrew University of Jerusalem, Jerusalem 91904, Israel.\and
    \small$^{3}$The Racah Institute of Physics, The Hebrew University of Jerusalem, Jerusalem 91904, Israel.\and
	\small$^\ast$Corresponding author. Email: ittai.shalev@mail.huji.ac.il
}

\begin{document}
\maketitle

\begin{abstract} \bfseries \boldmath
Recent advances in chip-scale magnetic quantum sensing have produced platforms that pair unprecedented sensitivity with extreme miniaturization. Here, we demonstrate a hybrid quantum sensor by combining Nitrogen-Vacancy (NV) centers in diamond with a rubidium (Rb) vapor cell, designed for precise magnetic field measurements and quantum exploration. The hybrid comagnetometer leverages the high resolution vector magnetic sensing of NV centers along with the high scalar-field sensitivity of the Rb vapor, enhancing the estimation of the magnetic field in terms of magnitude, direction and spatial distribution. A micromachined mm-scale vapor cell containing Rb atoms is paired with a bulk diamond, enabling optical and microwave control of both quantum systems for integrated field estimation. Simulations and experimental results confirm the improved accuracy of the system in magnetic field measurements, demonstrating a beyond 10 dB improvement. This NV–Rb platform offers a versatile route toward portable, sensitive magnetometry and opens new possibilities for integrated, multi-modal quantum sensing.

\end{abstract}

\maketitle

\noindent

Magnetic field sensors span a broad range of technologies, each offering distinct trade-offs between sensitivity, spatial resolution, bandwidth, and accuracy ~\cite{Budker2007, Degen2017, Hall2014_ReviewMagnetometry}. Quantum magnetometers, in particular, have emerged as powerful tools capable of pushing the limits of precision sensing across fundamental research and applied technologies ~\cite{Doherty2013, Rondin2014, Taylor2008}.

One prominent example is the nitrogen vacancy (NV) center in diamond, a well-established platform for quantum magnetometry \cite{Doherty2013, Rondin2014, Taylor2008}. NV centers are point defects in the diamond lattice consisting of a nitrogen atom adjacent to a vacancy. Their spin-dependent fluorescence allows highly sensitive, room-temperature magnetic field measurements, with sensitivities ranging from $\mu T/\sqrt{\mathrm{Hz}}$ down to $pT/\sqrt{\mathrm{Hz}}$ \cite{Taylor2008, Maze2008, Strner2021, Barry2024, Wolf2015}. NV centers measure the vector components of the magnetic field along the axis of each defect orientation; with four possible orientations in the lattice, measurements from three orientations are sufficient to reconstruct a full three-dimensional vector field \cite{Barry2020}. Although NV centers offer atomic-scale spatial accuracy in theory, practical spatial resolution is typically limited by optical diffraction to a few hundred nanometers or less \cite{Grinolds2013_NVdiffraction} — still exceptional among room-temperature sensors.

In contrast, alkali-atom magnetometers, such as those based on rubidium (Rb) vapor, operate by optically pumping atomic spins and measuring their response to magnetic fields via spin precession \cite{Budker2007, Kominis2003, Seltzer2008}. These sensors offer state-of-the-art sensitivities ranging from $nT/\sqrt{\mathrm{Hz}}$ to below $fT/\sqrt{\mathrm{Hz}}$ \cite{Budker2007, Dang2010}. However, their spatial resolution is typically limited by the physical dimensions of the vapor cell. While miniaturization can improve spatial resolution, it generally comes at the cost of sensitivity \cite{ Knappe2005_MicrofabricatedCells, Shah2007}. Additionally, Rb vapor magnetometers primarily measure the scalar magnitude of the magnetic field. Although vector magnetometry is possible using multi-axial configurations \cite{Sheng2013_RbVector}, these techniques are more complex and susceptible to alignment errors. Moreover, Bell–Bloom–based scalar magnetometers exhibit "dead zones" of maximal uncertainty along the light propagation axis, limiting directional sensitivity \cite{Budker2007, Bell1961}.

These complementary strengths and limitations motivate the development of hybrid sensing architectures \cite{Wolf2021_HybridSensing, Strner2021}. NV centers provide robust vector information with high spatial resolution, while Rb vapor cells offer high sensitivity in measuring the field’s magnitude \cite{Taylor2008, Grinolds2013_NVdiffraction, Budker2007}. However, neither system alone provides optimal performance across all sensing dimensions.

To address this, we developed an integrated hybrid quantum system that combines both NV-based and Rb-based sensing modalities. By placing an NV-dense diamond in direct contact with a micromachined millimeter-scale cell containing Rb vapor, we achieved a compact platform capable of simultaneous magnetic field measurements. In this configuration, the Rb system primarily measures the magnetic field magnitude, while the NV centers provide directional information.

We show that by combining the complementary data from both sensors, we can significantly reduce the uncertainty in the estimation of the magnetic field vector. Through a combination of experimental measurements and numerical simulations, we developed and validated a vector field estimator that improves magnitude sensitivity over NV-only sensing while preserving angular accuracy. This work presents a novel approach to hybrid quantum sensing and lays the foundation for future integrated systems capable of measuring multiple physical quantities with high spatial and temporal resolution.

\maketitle

\section*{Combined Magnetometry - schemes and simulations}

The designed hybrid quantum magnetometer combines Nitrogen-Vacancy (NV) centers in diamond and Rubidium (Rb) vapor, harnessing complementary quantum properties to enhance magnetic field detection \cite{Degen2017, Budker2007, BarGill2021, Stern2023, Pham2011}. This sensor’s architecture is designed for high-precision measurements, with a specific optical configuration and a tailored laboratory setup to support high sensitivity. The magnetic measurement protocol integrates data from both NV and Rb systems, providing a robust methodology for precise magnetic field estimation. Simulations further demonstrate the gains in sensitivity and accuracy achieved by this dual-system approach, offering significant advantages over standalone NV or Rb sensors.

\subsection*{Hybrid Sensor and Setup Configuration}

\begin{figure}
    \centering
    
    \includegraphics[width=0.6\textwidth]{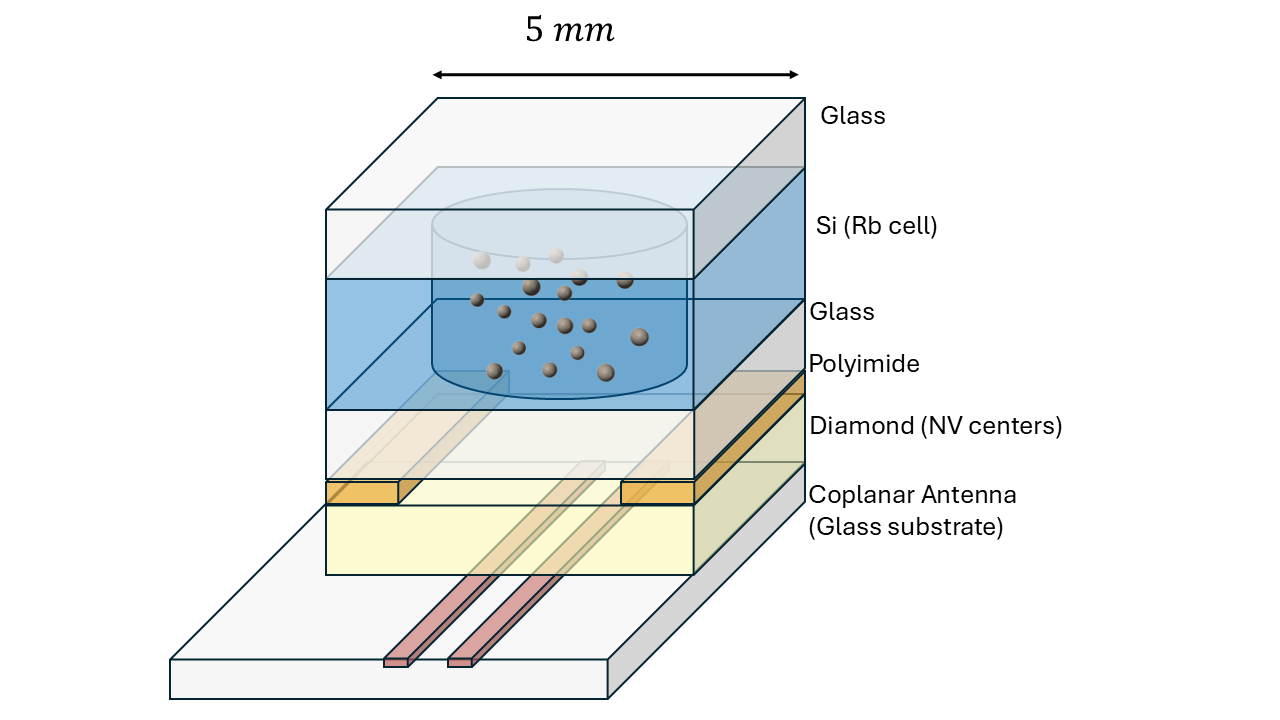}
    \begin{picture}(0,0)
        \put(0,0){\textbf{(a)}} 
    \end{picture}
    \hfill

    \includegraphics[width=0.6\textwidth]{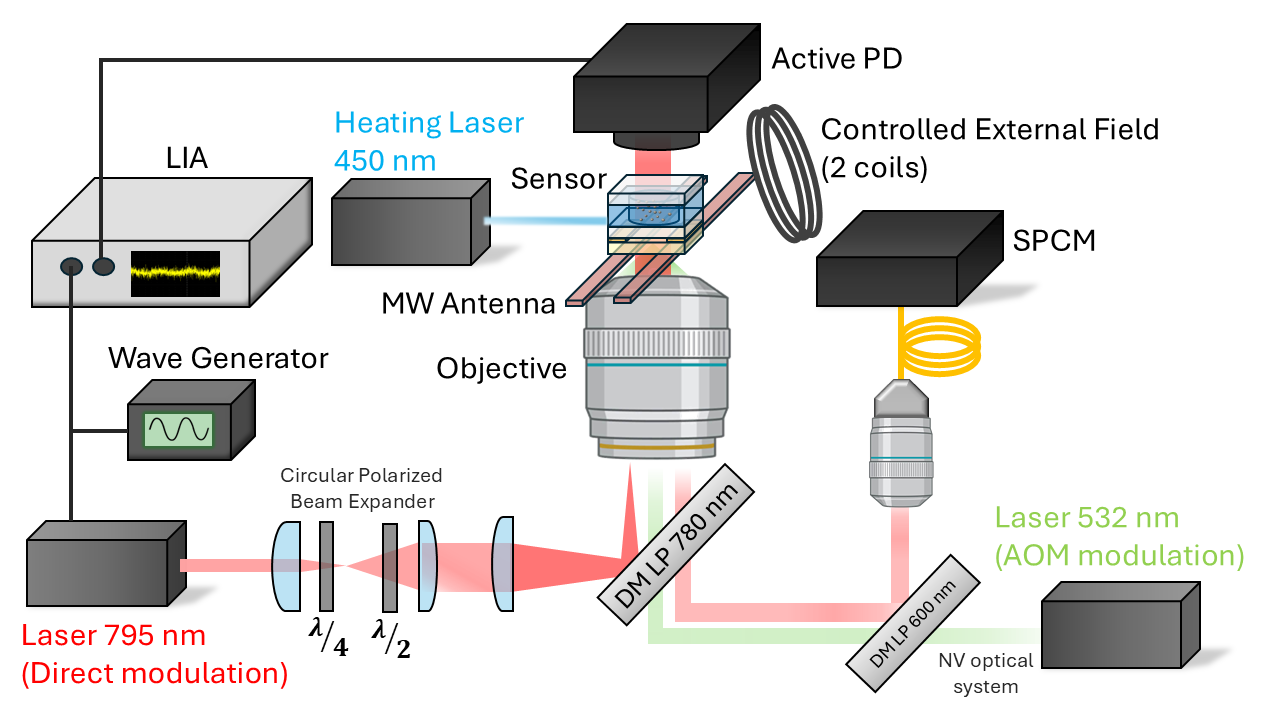}
    \begin{picture}(0,0)
        \put(0,0){\textbf{(b)}}
    \end{picture}
    \hfill

    \includegraphics[width=0.6\textwidth]{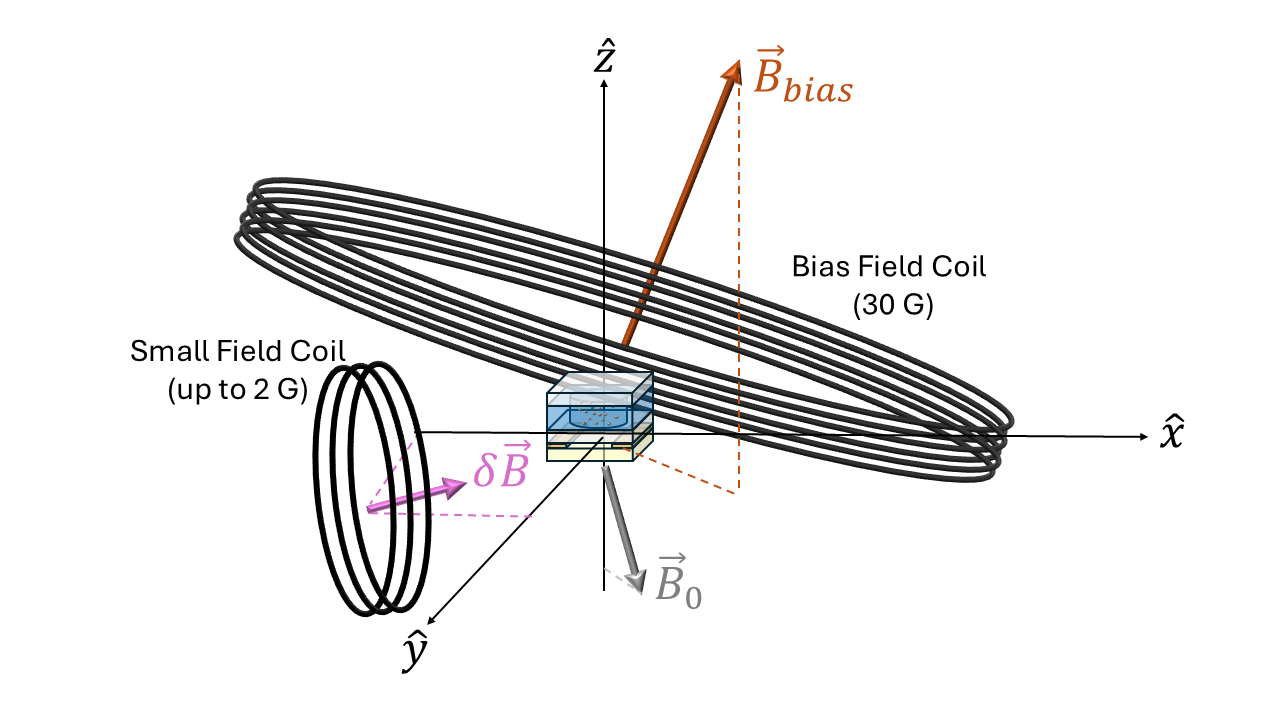}
    \begin{picture}(0,0)
        \put(0,0){\textbf{(c)}}
    \end{picture}
    \hfill
    
    \caption{\textbf{System and sensor configurations.}
    (\textbf{a}) Schematic of the integrated sensor utilized in the integrated magnetometer system.
    (\textbf{b}) Schematic of the integrated optical system of the magnetic sensor, illustrating the combined optical pathways of the Rb system and part of the NV system. This setup demonstrates how the optical components for both sensing mechanisms are integrated into a single configuration.
    (\textbf{c}) Configuration of the magnetic fields of the set-up of the integrated system.}
    \label{fig:combined_sensor}
\end{figure}
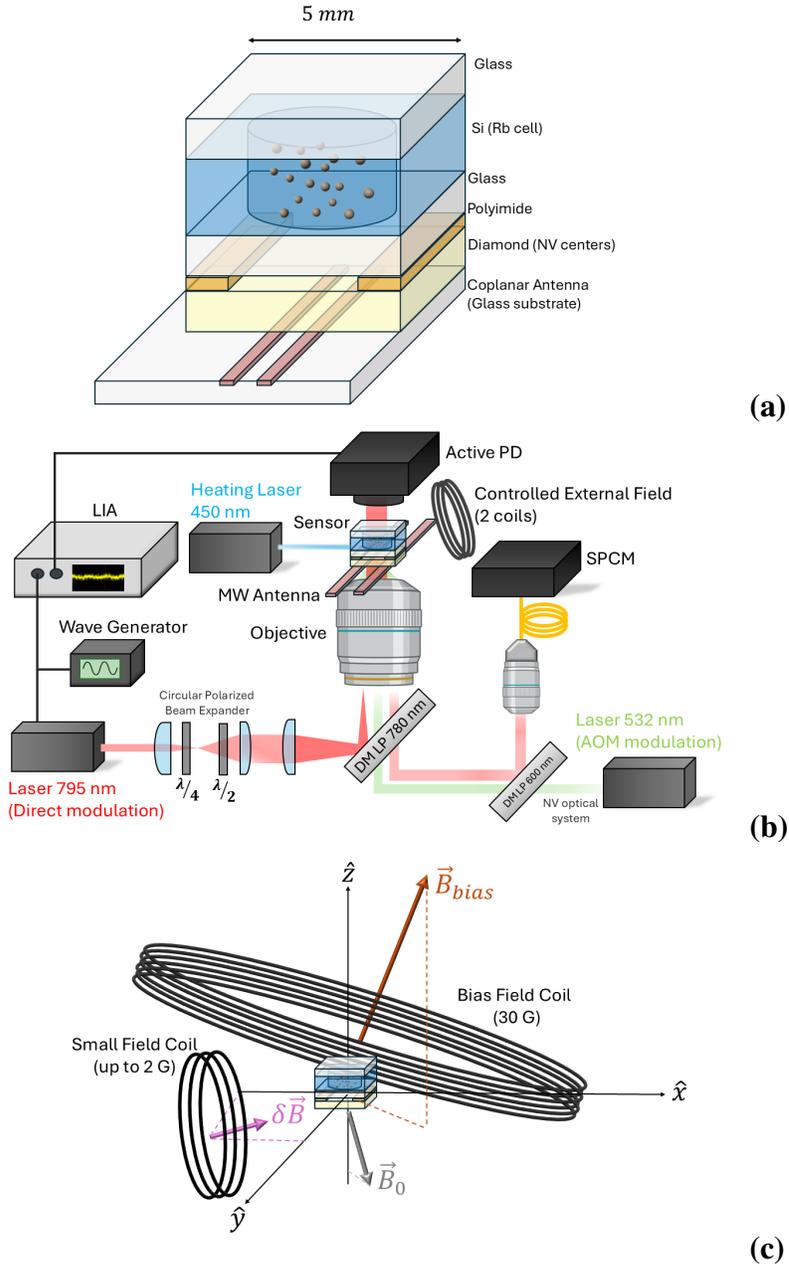

The main component of the combined setup is the integrated sensor, which facilitates magnetic sensing at a specific location and time. This integrated sensor comprises a mm-scale micromachined Rb cell positioned adjacent to a bulk diamond, with polyimide spacers separating the two components. Fig. \ref{fig:combined_sensor} (A) illustrates the structure of the integrated sensor.

The experimental setup [Fig.~\ref{fig:combined_sensor} (B)] integrates a hybrid quantum sensor that combines nitrogen-vacancy (NV) centers in diamond with rubidium (Rb) vapor for magnetic field measurements. The NV subsystem operates using optically detected magnetic resonance (ODMR~\cite{Gruber1997}, while the Rb subsystem employs Bell–Bloom magnetometry~\cite{Dang2010}, using a single modulated laser beam for both optical pumping and probing.

Magnetic field control is achieved using two independent coils [Fig.~\ref{fig:combined_sensor} (C)]. A large Helmholtz coil with a 12 cm radius and 500 turns surrounds the sensor and generates a static bias field $B_{\text{bias}}$ of approximately 30 G when driven with a 1 A current. This field lifts the degeneracy of the NV spin levels and sets the quantization axis. A smaller coil, 2.54 cm in radius with 40 turns, is positioned near the sensor and generates a tunable measurement field $\delta B$ up to 2 G, controlled via currents up to 4 A. In addition to these applied fields, a static background field $B_0$, primarily due to the Earth’s magnetic field, is present in the environment. These field components jointly determine the total magnetic environment experienced by the sensor and are crucial for accurate vector and scalar field reconstruction.

\subsection*{DC Magnetic Field Measurement Method and Sensitivity Calculation}

\subsubsection*{NV Center Measurement}

The NV centers system performs an ODMR scan using the MW antenna, resulting in a scan that represents the projection of the magnetic field along each of the orientations \( a \), \( b \), \( c \), and \( d \) according to the Larmor precession relation \(B = \frac{2\pi f}{\gamma}\) \cite{Jelezko2006, Taylor2008, Maze2008}. Such a measurement, with a known bias field $B_{\text{bias}}$ and given the diamond's crystallographic structure, allows for the extraction of the vector magnetic field of interest.

\begin{figure}[tbh]
    \centering
    \includegraphics[width = 1 \linewidth]{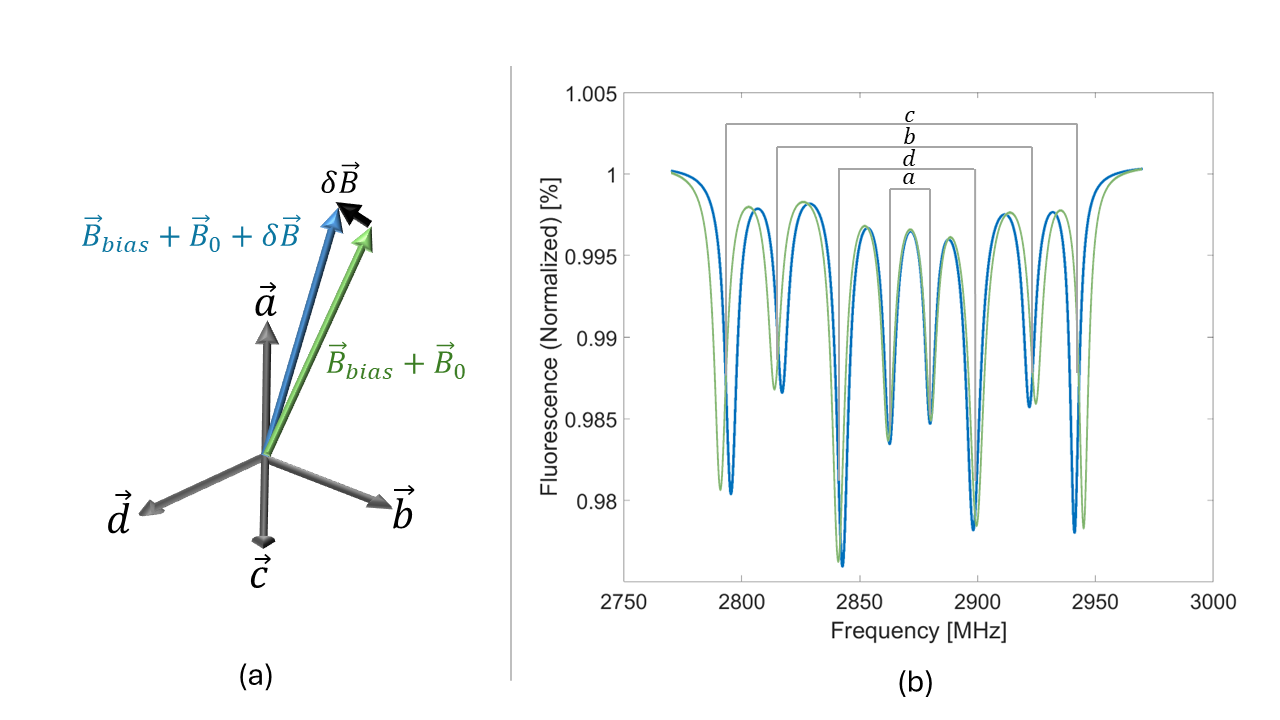}
    \caption{\textbf{Magnetic field sensing using NV centers}.  \textbf{(a)} A diagram illustrating the method for extracting the desired magnetic field as the difference between two distinct measurements obtained from the NV system. \textbf{(b)} Two separate ODMR measurements are depicted: In blue, the measurement with the small magnetic field activated, and in green, the measurement without the activation of the small field. By subtracting the green measurement from the blue measurement, the desired small magnetic field can be accurately extracted.}
    \label{fig:NV_measurement_visualisation}
\end{figure}

It is important to note that, in practice, we aim to measure an unknown magnetic field \(\delta \mathbf{B}\). This can be achieved through two ODMR scans: the first scan is conducted with the addition of the small field \(\delta \mathbf{B}\) alongside the bias field \(\mathbf{B}_{\text{bias}}\), while the second measurement is performed without applying the small field. Subsequently, by subtracting the results of the two scans, we can isolate and determine the desired small field (Fig. \ref{fig:NV_measurement_visualisation}):

\begin{equation}
\delta \mathbf{B} = \left( \mathbf{B}_{\text{bias}} + \delta \mathbf{B} + \mathbf{B}_0 \right) - \left( \mathbf{B}_{\text{bias}} + \mathbf{B}_0 \right)
\label{eq:delta_B}
\end{equation}

Note that a background field \( \mathbf{B}_0 \) is included in both measurements, representing the background magnetic field present in the system. In a shielded system, this background field may be negligible; however, in an unshielded system, the background field can be significant in relation to the small field \( \delta \mathbf{B} \) \cite{Doherty2013, Rondin2014}.

The empirical sensitivity of the NV measurement is derived from the maximum slope \( m_{\text{NV}} \) observed at each peak of the ODMR spectrum. This sensitivity is influenced by the empirical uncertainty \( \Delta PL \) of the photoluminescence (PL) measurement at the frequency \( f_{\text{max}} \) where the slope is at its maximum. Additionally, it incorporates the gyromagnetic ratio \( \gamma_{\text{NV}} \) of the NV. The empirical sensitivity related to a specific ODMR orientation can be expressed as follows \cite{Taylor2008, Barry2020}:

\begin{equation}
\Delta {B}'_{\text{NV}} = \frac{\Delta f}{\gamma_{\text{NV}}} = \frac{\Delta PL}{\gamma_{\text{NV}} m_{\text{NV}}}
\label{eq:delta_B_NV}
\end{equation}

Each frequency in the ODMR measurement is associated with an exposure time $t_m$, which directly impacts the uncertainty in the PL signal \cite{Wolf2015}. As the detector's exposure time increases, the uncertainty in the PL measurement decreases. Consequently, this leads to an enhancement in the accuracy of the magnetic field measurement, resulting in a reduction of $\Delta {B}'_{\text{NV}}$.

\subsubsection*{Rb Vapor Cell Measurement}

The measurement of the magnetic field using Rb is conducted by scanning the modulation frequencies of the modulated laser and identifying the resonance in the spin frequency response of the cell, which is reflected in the output signal of the LIA. The resonance corresponds to the peak value of the in-phase \(X\) component of the LIA signal or the \(R\) component representing the magnitude of the signal. Alternatively, it is possible to estimate the resonance frequency by locating the point of maximum slope in the out-of-phase \(Y\) component of the LIA signal~\cite{Seltzer2008, Bell1961}.

It is important to note that the field measured using Rb is, in fact, the scalar field \(|\delta \mathbf{B} + \mathbf{B}_0|\), which incorporates the contributions of the background field.

The uncertainty of the Rb measurement is determined through a linear approximation at the point of maximum slope of the Y signal~\cite{Scofield1994}. In this region, the maximum slope \(m_{\text{Rb}}\) is calculated, and the root mean square error (RMSE) of the Y signal is found using a linear fit (LS fit) in the linear approximation region~\cite{Bevington2003}. The RMSE effectively represents the empirical uncertainty of the measurement, denoted as \(\Delta Y\).

Consequently, we can derive the uncertainty associated with the magnetic field measured using the Rb system:

\begin{equation}
\Delta B_{\text{Rb}} = \frac{\Delta f}{\gamma_{\text{Rb}}} = \frac{\Delta Y}{\gamma_{\text{Rb}} m_{\text{Rb}}}
\label{eq:delta_B_Rb}
\end{equation}

\subsubsection*{DC Magnetic Field Estimation Using Combined System}

Angular information regarding the measured magnetic field \(\delta \mathbf{B}\) is exclusively obtained from the magnetic field measurement of the NV system, denoted as \(\mathbf{B}_{\text{NV}}^{(\delta \mathbf{B})}\) \cite{Pham2011, Barry2020}. In contrast, the rubidium measurement, represented as \(B_{\text{Rb}}^{(\delta \mathbf{B})}\), provides a high-precision measurement of the magnitude of the magnetic field \(\delta \mathbf{B}\) but yields significant uncertainty concerning the direction of the field \cite{Seltzer2008, Budker2007}. Therefore, we aim to construct an estimator for the magnetic field given by:

\begin{equation}
\mathbf {\hat B}_{{\delta B}} = \mathbf{B}_{\text{NV}}^{(\delta B)} - \mathbf{\hat B}_{{c}}
\label{eq:B_deltaB}
\end{equation}

To determine the correction vector \( \mathbf{\hat B}_{{c}} \), we seek the minimal adjustment that ensures the combined magnitude of the estimated field, along with the background field \( \mathbf{B}_0 \), equals the field magnitude measured by the Rb system, \( B_{\text{Rb}}^{(\delta B)} \). This approach allows for a minimal modification of the NV measurement while maintaining angular accuracy, ensuring that the magnitude of the estimator matches the Rb measurement \( B_{\text{Rb}}^{(\delta B)} \). The correction \( \mathbf{\hat B}_{{c}} \) is determined by solving the following optimization problem:

\begin{equation}
\arg \min_{\mathbf{\hat B}_{{c}}} (\|\mathbf{B}_{\text{NV}}^{(\delta B)} + \mathbf{B}_0 - \mathbf{\hat B}_{{c}} \|) = B_{\text{Rb}}^{(\delta B)}
\label{eq:optimization_problem}
\end{equation}

According to Eq. \ref{eq:optimization_problem}, we observe that in the correction space of \( \mathbf{\hat B}_{{c}} \), we essentially obtain a sphere centered at the point \( \mathbf{B}_{\text{NV}}^{(\delta B)} + \mathbf{B}_0 \), with a radius defined by \( B_{\text{Rb}}^{(\delta B)} \). Our objective is to identify the point on the surface of this sphere that is closest to the origin of the coordinate axes. This point represents the solution that satisfies the minimum requirement of Eq. \ref{eq:optimization_problem}, and is given by the following analytical expression:

\begin{equation}
\mathbf{\hat B}_{{c}} = \left( \frac{\mathbf{B}_{\text{NV}}^{(\delta B)} + \mathbf{B}_0}{\|\mathbf{B}_{\text{NV}}^{(\delta B)} + \mathbf{B}_0 \|} \right) (\|\mathbf{B}_{\text{NV}}^{(\delta B)} + \mathbf{B}_0 \| - B_{\text{Rb}}^{(\delta B)})
\label{eq:optimization_solution}
\end{equation}

Under the assumption that the background field is significantly smaller than the small field of interest, \( \mathbf{B}_0 \ll \delta \mathbf{B} \), we can neglect the background field and estimate the magnetic field as follows:

\begin{equation}
\arg \min_{\mathbf{\hat B}_{{c}}} (\|\mathbf{B}_{\text{NV}}^{(\delta B)} - \mathbf{\hat B}_{{c}} \|) = B_{\text{Rb}}^{(\delta B)}
\label{eq:optimization_problem_no_B0}
\end{equation}

\begin{figure}[tbh]
    \centering
    \includegraphics[width = 0.9 \linewidth]{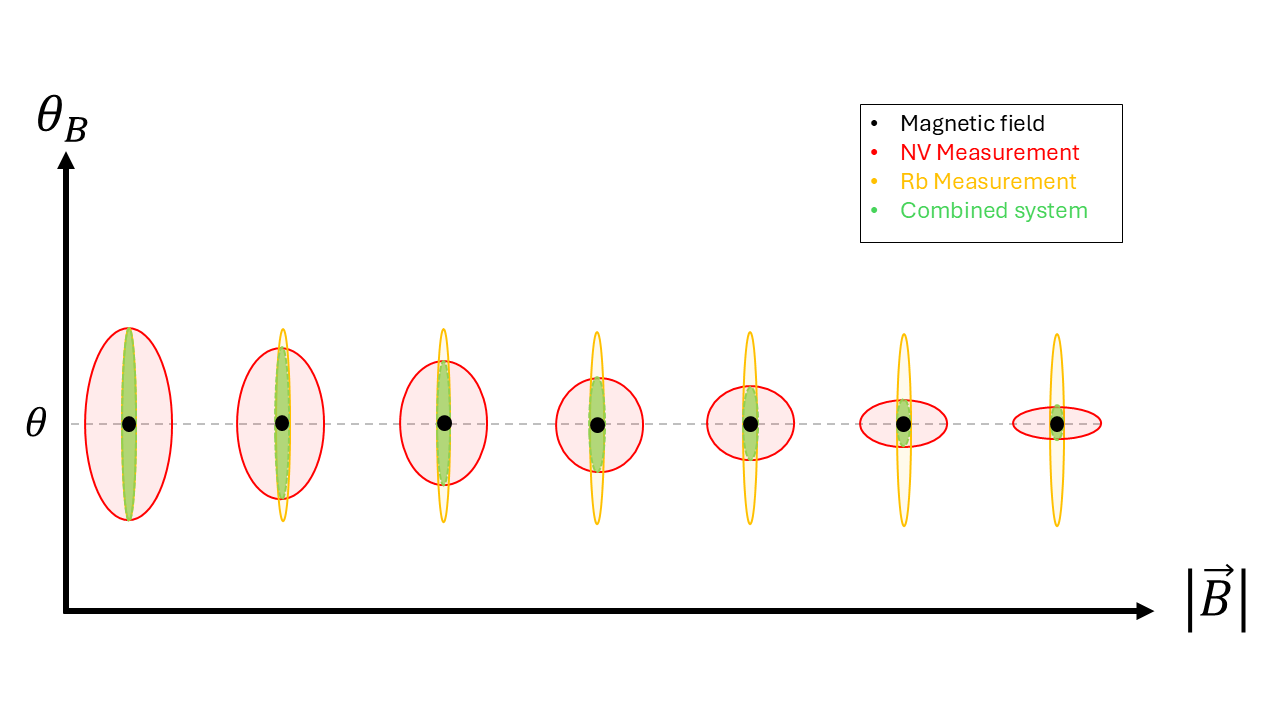}
    \caption{\textbf{Illustration of magnetic field magnitude and orientation estimation employing both the NV and Rb systems, along with the combined estimator.} This representation includes the associated errors depicted in a polar coordinate system. As the magnitude of the magnetic field increases, the angular accuracy of the NV system also increases, so as the accuracy of the combined estimation (more details in Supplementary Text - Angular accuracy). On the other hand, the accuracy of the magnetic field's magnitude is constant and determined by the Rb measurement.}
    \label{fig:all_measurements_scheme}
\end{figure}

The combined estimator demonstrates a dual advantage: it fully leverages the angular uncertainty provided by the NV measurement which is much lower than the Rb scalar measurement, and conversely, the overall accuracy of the estimator's magnitude is primarily dictated by the precise rubidium measurement which has a much lower uncertainty than that of the NV system \cite{Glenn2018}. This interplay of strengths is illustrated in Fig. \ref{fig:all_measurements_scheme}.

In the case where the background field is significantly smaller than the small field of interest, the improvement in the magnitude uncertainty of the combined estimation, compared to the NV system alone, will be given by the ratio \(\sigma_{\text{NV}} / \sigma_{\text{Rb}}\), which is typically on the order of three magnitudes. On the other hand, the combined system provides angular estimation corresponding to the NV system measurements, thereby enhancing and supplying angular information that the Rb measurement alone cannot provide. 'Supplementary Text - Shielded Environment Measurement Simulation' showcases the performance of a combined estimation using a simulation.

\subsubsection*{Effect of Unshielded Environment}

Up to this point, we have focused on the case where the background field \(\mathbf{B}_0\) is negligible. Now, we will examine the scenario in which \(\mathbf{B}_0\) is encountered when measurements are taken in an unshielded environment. The challenge here lies in the fact that while the Rb measurement is sensitive to the background field, modeled as \(|\delta \mathbf{B} + \mathbf{B}_0|\), the NV measurement remains unaffected by \(\mathbf{B}_0\) as it is canceled out, as shown in Eq. \ref{eq:delta_B}. Therefore, a separate estimation of \(\mathbf{B}_0\) is required.

To estimate \(\mathbf{B}_0\), we will conduct three calibration measurements. In each calibration, a relatively strong field will be applied and measured using the combined system. For each \(i^{\text{th}}\) calibration measurement, the following condition must be satisfied:

\begin{equation}
\|\mathbf{B}_0 + \mathbf{B}_{\mathbf{NV}}^{(\text{cal},i)}\| = {B}_{{Rb}}^{(\text{cal},i)}
\label{eq:calibration_condition}
\end{equation}
where \( \mathbf{B}_{\mathbf{NV}}^{(\text{cal},i)} \) and \( {B}_{{Rb}}^{(\text{cal},i)} \) are the NV and Rb measurements of the \( i^{th} \) calibration field. Eq. \ref{eq:calibration_condition} represents a system of non-linear equations with three variables: \( B_{0x}, B_{0y}, B_{0z} \). By solving this system using a numerical method, an estimation for \( \mathbf{B}_0 \) can be obtained \cite{Budker2007,Shah2007}.

It is important to note that the NV measurement of the calibration fields introduces noise, with a magnitude comparable to the inherent uncertainty of the NV system \cite{Degen2017}, which can significantly affect the overall combined estimation. To improve the accuracy of the combined estimation in the presence of a background field, the calibration measurements for the NV system should be performed using ODMR over a longer period \(T\) (assuming that shot noise is the dominant noise in the system, in which the PL uncertainty will decrease by \(\sqrt{T}\)) \cite{Taylor2008}.

After estimating the background magnetic field, the combined estimation can be done using Eq. \ref{eq:optimization_solution}. To evaluate the combined estimation, a simulation that estimates the magnetic field using both the combined system and the NV system independently is presented. In this simulation, we defined the uncertainty of the NV system to be 1,000 times larger than that of the Rb system. The magnetic field was estimated within the range of -1.5 G to 1.5 G, and was constrained to the x-y plane, while noise was present in three dimensions. Each field value was tested through 50 repetitions to ensure sufficient statistical data. Subsequently, we calculated the improvement of the combined system's measurements compared to those obtained using the NV system alone, assessing both the magnitude and direction of the magnetic field. The improvement is quantified by the ratio of the mean square error (MSE) between the combined estimator and the NV-only measurement, expressed on a logarithmic scale (dB), for each simulated magnetic field.

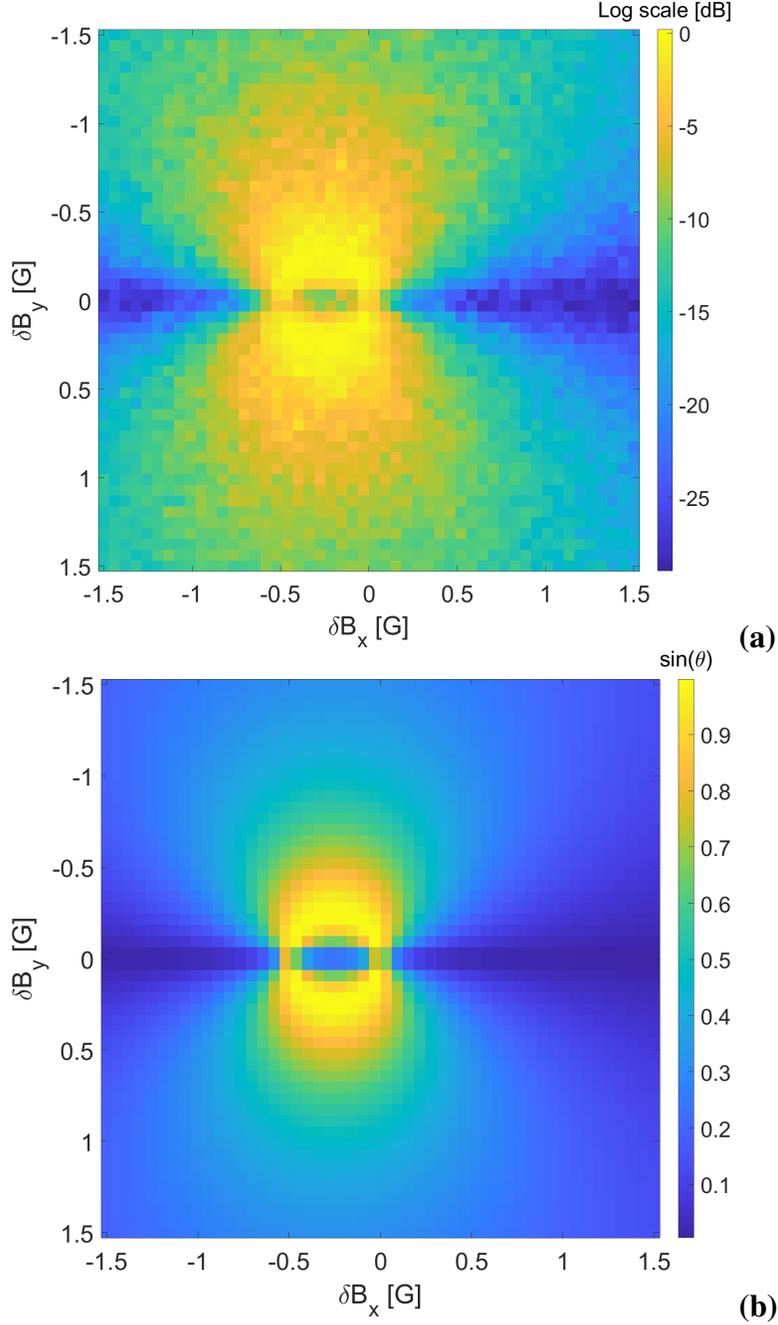
\begin{figure}
    \centering
    
    \includegraphics[width=0.58\textwidth]{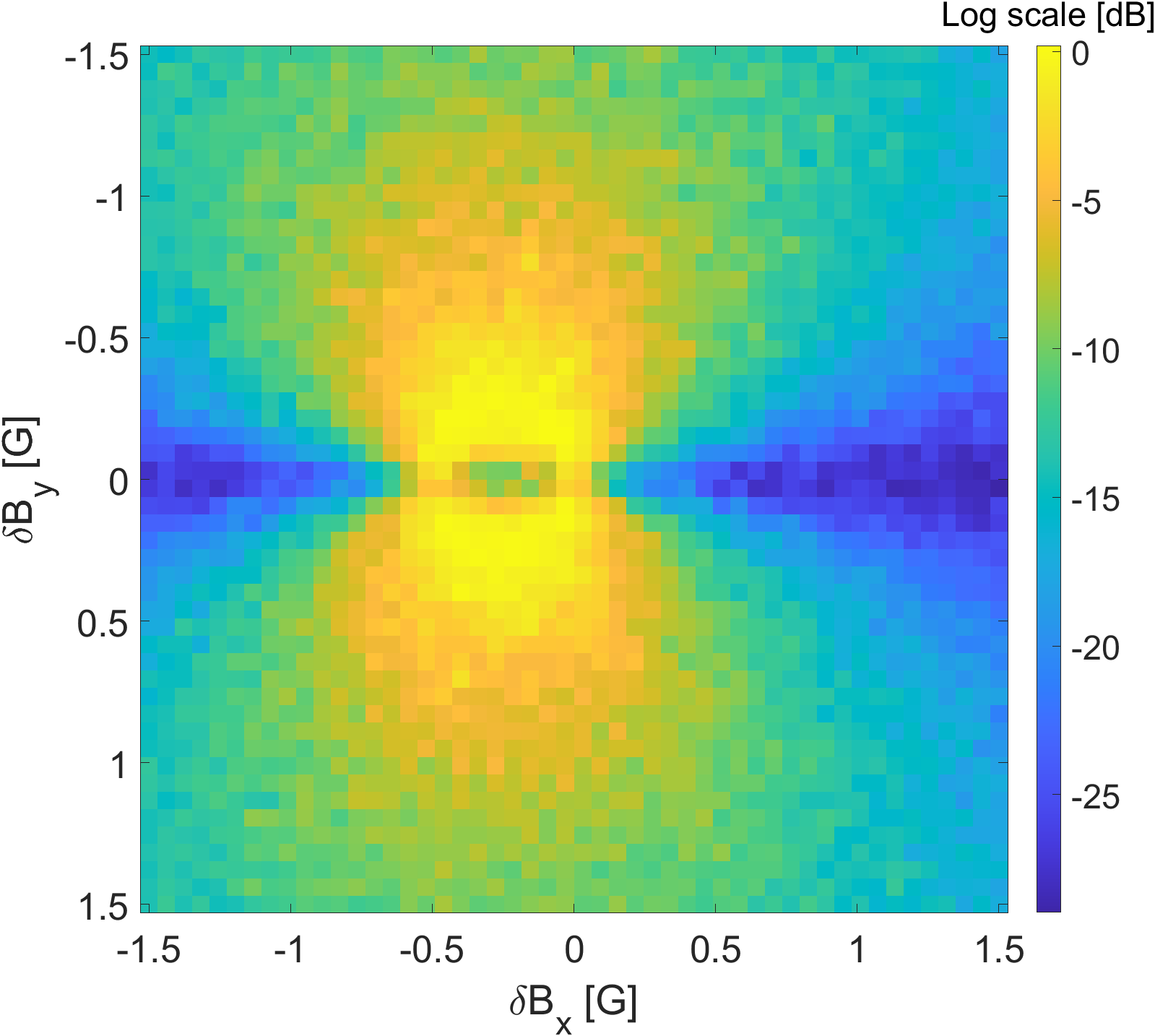}
    \begin{picture}(0,0)
        \put(0,0){\textbf{(a)}} 
    \end{picture}
    \hfill

    \includegraphics[width=0.58\textwidth]{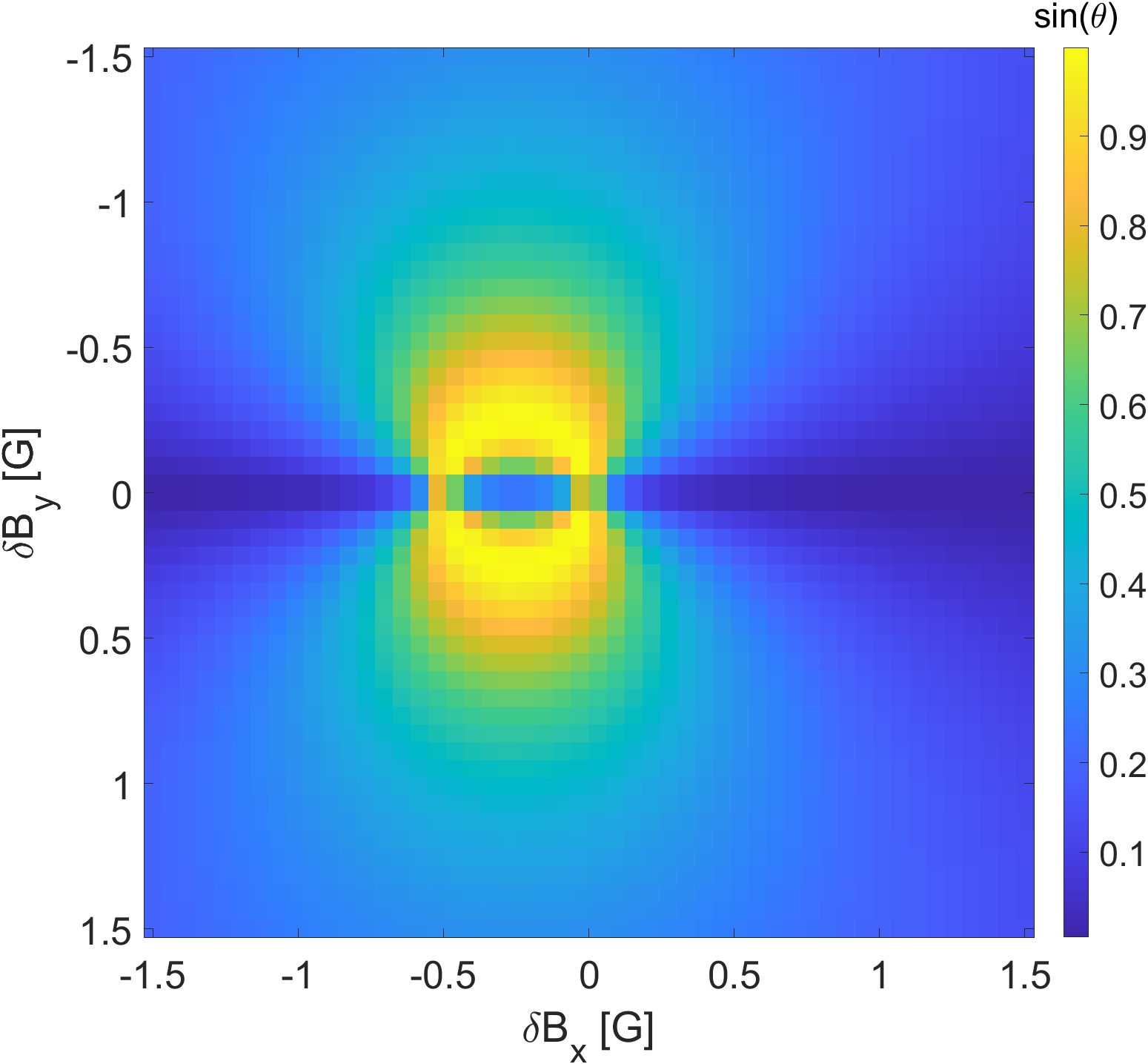}
    \begin{picture}(0,0)
        \put(0,0){\textbf{(b)}}
    \end{picture}
    \hfill
    
    \caption{\textbf{Combined estimation simulation.}
    (\textbf{a}) Simulation of the combined estimation in the presence of a background field \( \vec{B}_0 \) with a magnitude of 0.5 G at the \(x\) direction, illustrating the enhancement in estimating the magnetic field magnitude using the combined estimation approach compared to the NV system alone.
    (\textbf{b}) Simulation of the orthogonality of the correction \( \mathbf{\hat B}_{{c}} \) (which is parallel to $\vec{B}_{\text{NV}}^{(\delta B)} + \vec{B}_0$), relative to the measured field $\delta \vec{B}$ in the presence of a background field \( \vec{B}_0 \) with a magnitude of 0.5 G at the \(x\) direction. Therefore, this figure effectively demonstrates the orthogonality of \( \mathbf{\hat B}_{{c}} \) in relation to the NV measurement $\vec{B}_{\text{NV}}^{(\delta B)}$, excluding any measurement noise.
    }
    \label{fig:simulation_with_B0}
\end{figure}

Fig. \ref{fig:simulation_with_B0} (a) illustrates the effect of the background field on the improvement in estimation of the combined system compared to the NV system alone in terms of MSE (in logarithmic scale).

The simulations indicate that the maximum improvement in terms of MSE can reach up to 25 dB, taking into consideration the specific parameters of our experimental system (which are introduced in 'Materials and Methodods - Rubidium and NV measurements and sensitivities' \cite{methods}) as well as a specific uncertainty of the background field, which depends on the calibration duration.

Additionally, a pattern can be observed in which the performance of the combined estimator does not enhance the estimation at all, indicated by the measured fields where the improvement is 0 dB. This phenomenon arises from the specific design of our combined estimator.

Analytically, as shown in Eq. \ref{eq:optimization_solution}, the correction of the combined estimator \( \mathbf{\hat B}_{{c}} \) must be directed along $\vec{B}_\text{NV}^{(\delta B)} + \vec{B}_0$. For a small correction relative to the measured field $\vec{B}_\text{NV}^{(\delta B)}$, we can decompose the correction vector \( \mathbf{\hat B}_{{c}} \) into two orthogonal components \({\hat B}_{{r}} \) and \({\hat B}_{{\theta}} \). The first-order correction reveals that \({\hat B}_{{r}} \) effectively stretches the estimation of $\vec{B}_\text{NV}^{(\delta B)}$, while \({\hat B}_{{\theta}} \) induces a rotation in the estimation.

For the case that rotation component \({\hat B}_{{\theta}} \) is more dominant, the combined estimator will effectively rotate the NV estimation, resulting in minimal stretching. Conversely, when the stretching component \({\hat B}_{{r}} \) is dominant, the combined estimator will have little effect on rotating the NV estimation and will primarily stretch it. As previously noted, the primary goal of the combined estimator is to enhance the estimation of the magnetic field magnitude; thus, a rotational adjustment will be ineffective and yield no improvement, in contrast to a stretching adjustment, which will provide significant enhancement.

Fig. \ref{fig:simulation_with_B0} (b) presents the orthogonality analysis, where plot (b) directly corresponds to the patterns observed in plot (a) (in logarithmic scale), explaining the unique behavior and results obtained in the combined estimation. 'Supplementary Text - Enhanced Measurement At The Presence Of Background Magnetic Field' demonstrates a technique that enhances the combined estimator even in the presence of a significant background field.

\section*{Experiments and Results}

\subsection*{Combined Estimation Uncertainty}

\begin{figure}
    \centering
    \includegraphics[width=0.9\linewidth]{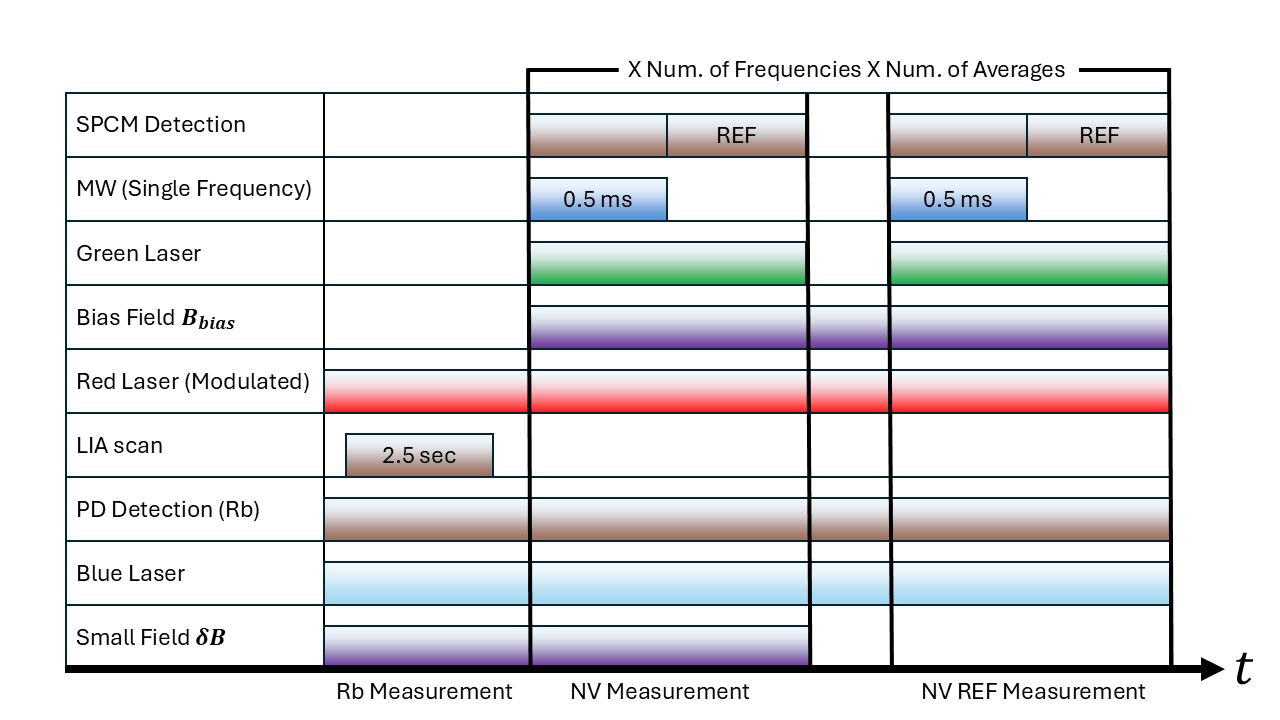}
    \caption{\textbf{Schematic representation of the signal timing during a single estimation cycle of the combined system, conducted over $N$ experiments.}}
    \label{fig:sequence}
\end{figure}

In the previous chapter, we presented two-dimensional simulations of the combined estimator's performance. In the experiment, we aim to demonstrate the performance of the combined estimator in one dimension by measuring a variable DC magnetic field along a single axis. This can be interpreted as measuring the marginal distribution of the combined estimator's performance improvement along one axis, with the fields along the perpendicular axes set to 0 G.

In the experiment, a small magnetic field $\delta \vec{B}$ is applied, as described in the Setup Configuration section. This field is first measured using the Rb system, followed by an immediate measurement with the NV system. This measurement sequence is repeated $N$ times for each small $\delta \vec{B}$. Fig. \ref{fig:sequence} illustrates the timing and sequence of the experiment, showing the various fields and laser operations used in each common measurement.

\begin{figure}
    \centering
    
    \includegraphics[width=0.9\textwidth]{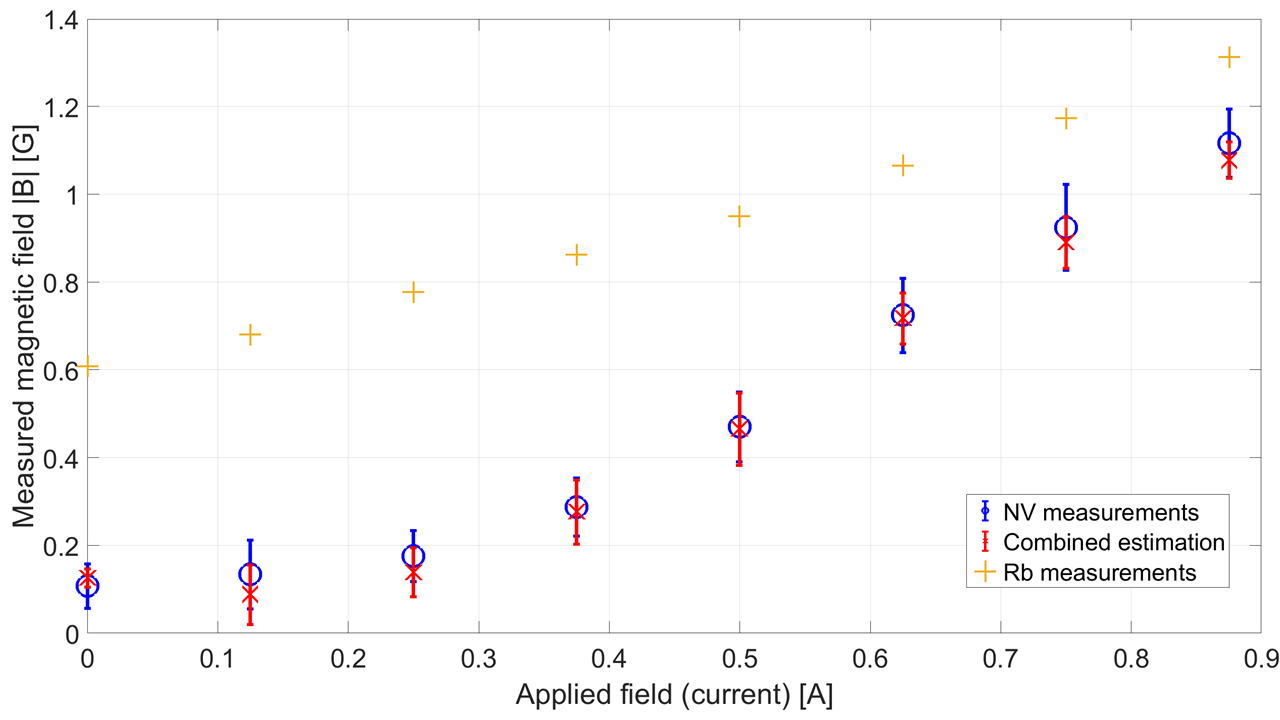}
    \begin{picture}(0,0)
        \put(0,0){\textbf{(a)}} 
    \end{picture}
    \hfill

    \includegraphics[width=0.9\textwidth]{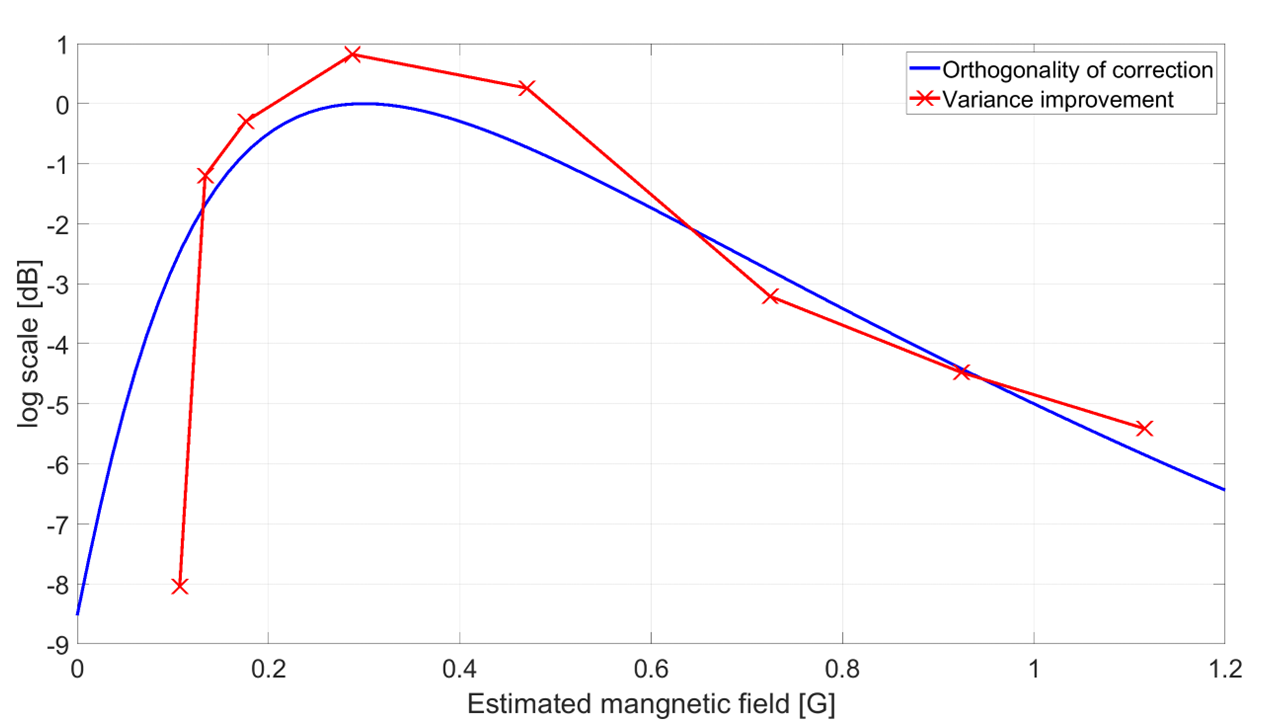}
    \begin{picture}(0,0)
        \put(0,0){\textbf{(b)}}
    \end{picture}
    \hfill
    
    \caption{\textbf{Combined estimation measurements.}
    (\textbf{a}) Average magnetic field measurements ($N=50$) of the combined system, along with separate Rb and NV measurements, including their associated uncertainties.
    (\textbf{b}) Improvement of the variance of the combined estimator relative to NV measurements alone, shown for both simulations and experimental data.
    }
    \label{fig:distribution_field_measurements}
\end{figure}
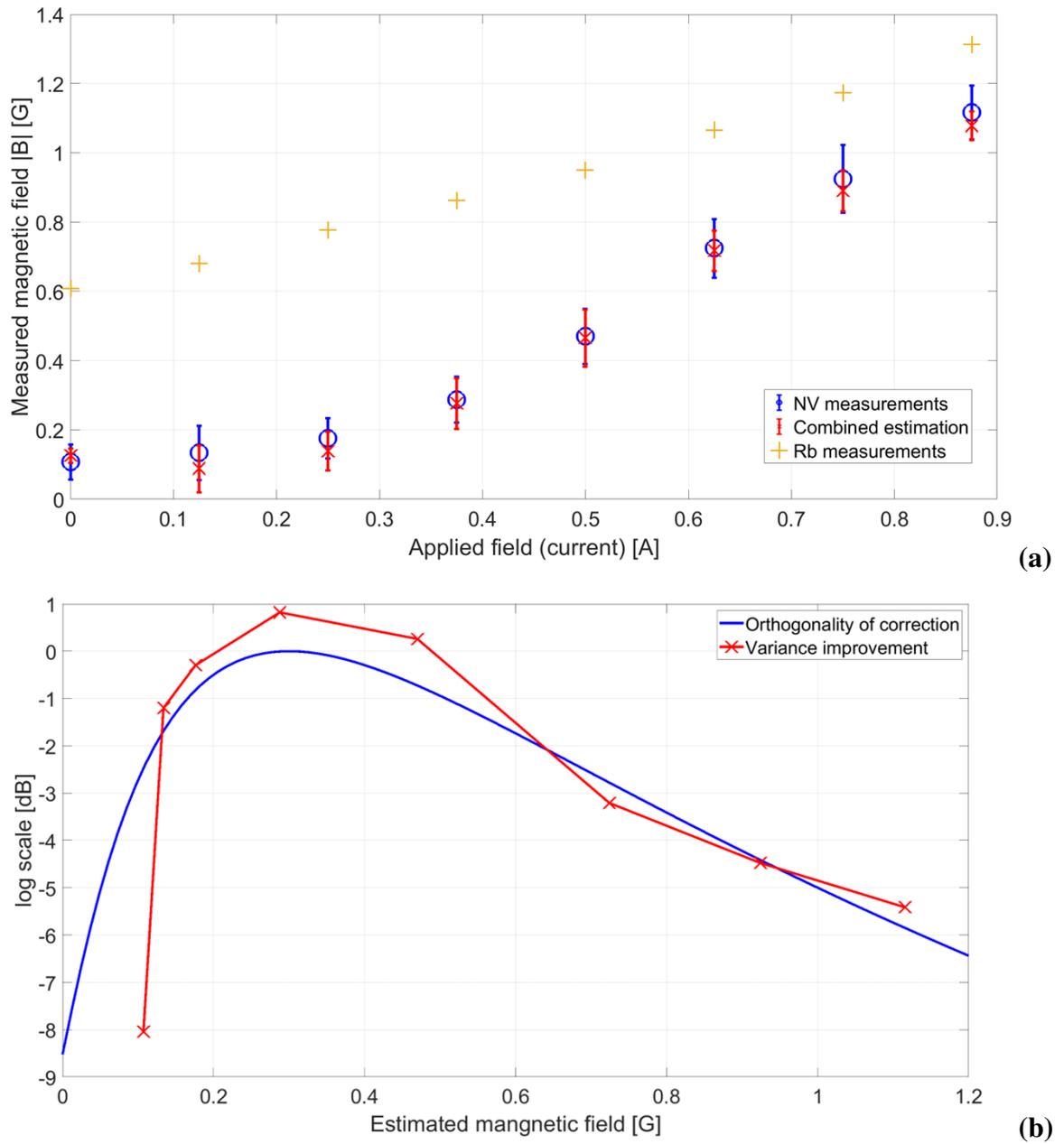

Following the methods presented earlier, the magnetic field was measured and extracted by each system within the combined setup using the least squares (LS) fitting method, with the RMSE of the measurements corresponding to the uncertainty of each system. The measurement results are shown in Fig. \ref{fig:distribution_field_measurements} (a).

Since the experimental setup operates outside a magnetically shielded environment, background magnetic fields significantly influence the measurement and must be carefully accounted for, as described in Eq.~\ref{eq:calibration_condition}. By empirically estimating the background field and the applied small fields along with their uncertainties, we simulated the marginal distribution of the combined estimator’s improvement. Additionally, the orthogonality between the correction vector \( \mathbf{\hat B}_{{c}} \) and the desired small field \( \mathbf{\delta B} \) was simulated to evaluate its impact on performance. Fig. \ref{fig:distribution_field_measurements} (b) shows both the improvement of the combined system relative to the NV system alone and the correlation between measured improvement and theoretical predictions based on this orthogonality condition.

These experimental results closely match the simulations, thus validating the theoretical analysis and demonstrating enhanced magnetic field estimation using a hybrid Rb-NV magnetometer system.

\subsection*{Field Distribution Measurement}

After obtaining the marginal distribution of the enhanced magnetic field estimation as previously described, we aimed to measure the magnetic field distribution of a small paramagnet using our combined system. To achieve this, we incorporated a linear stage into the setup, positioning a small magnet at its tip. This enabled us to measure the magnetic field generated by the magnet at various spatial points. The range of the linear stage used in our experiment is 30 mm, which provides flexibility in positioning the paramagnet for various measurements.

\begin{figure}
    \centering
    
    \includegraphics[width=0.8\textwidth]{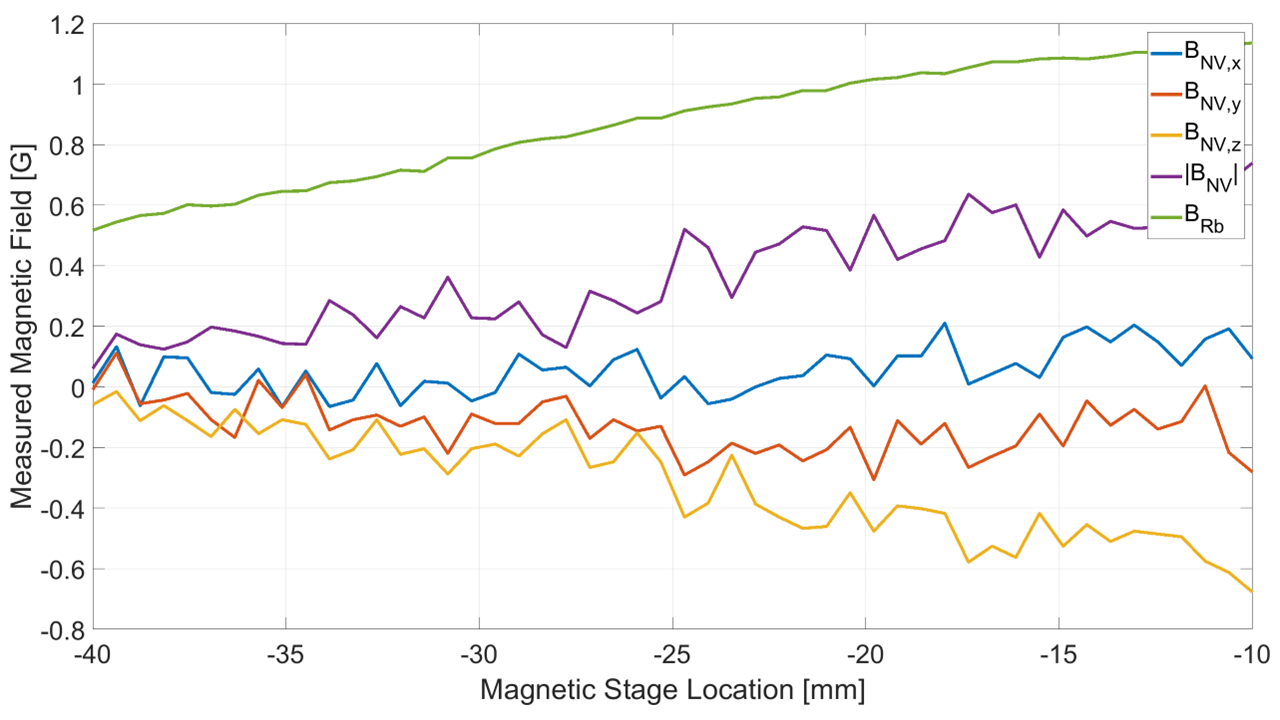}
    \begin{picture}(0,0)
        \put(0,0){\textbf{(a)}} 
    \end{picture}
    \hfill

    \includegraphics[width=0.8\textwidth]{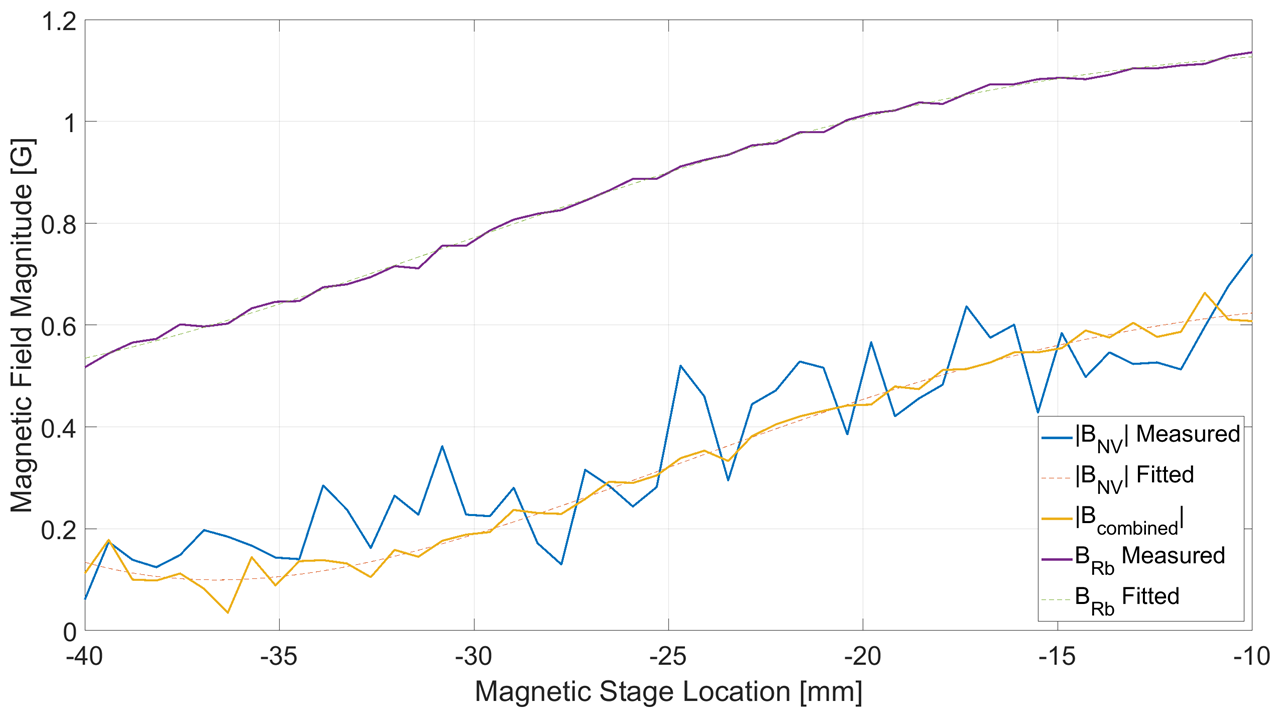}
    \begin{picture}(0,0)
        \put(0,0){\textbf{(b)}}
    \end{picture}
    \hfill
    
    \caption{
    \textbf{Combined estimation measurement of a spatial field at different locations.}
    (\textbf{a}) Magnetic field measurements performed separately by the NV system and the Rb system. The Rb system measures the scalar magnetic field, while the NV system provides the three components of the vector magnetic field. This contrast demonstrates the scalar precision of the Rb system, and the directional resolution offered by the NV system.
    (\textbf{b}) Improvement in magnetic field estimation. The purple line represents the Rb system measurement along with a second-degree polynomial fit (in red line). The blue line shows the NV system’s measurement of the magnetic field magnitude, while the yellow line is the combined estimation of the magnetic field magnitude. Additionally, the green line corresponds to the second-degree polynomial fit applied to the NV system's measurement. This figure highlights how the combined estimator enhances accuracy by incorporating both magnitude and directional information.
    }
    \label{fig:moving_field}
\end{figure}

In the experiment, we measured the magnetic field at 50 equally spaced positions along the range of the linear stage, using three different approaches: the Rb system, the NV system, and the combined magnetic field estimation. Each system provided independent measurements at these positions, allowing us to compare the performance of the individual sensors as well as the combined estimator.
Fig. \ref{fig:moving_field} (a) presents the magnetic field measurements conducted independently by the NV system and the Rb system. As anticipated, the influence of the background field $\vec{B}_0$ is significant in the Rb measurement, contributing approximately 0.5 G to the total measured field. In contrast, the NV system is insensitive to the background field $\vec{B}_0$.

Next, we applied the combined estimator to calculate the magnetic field. Fig. \ref{fig:moving_field} (b) presents the magnitude of the magnetic field as determined by the combined estimator, alongside the measurements from both the NV and Rb systems. It is very noticable that the combined estimator exhibits less noise than the NV system measurement, specifically, the root mean square error (RMSE) of the NV system, assuming the "true" magnetic field is represented by the polynomial fit (shown as the green curve), is 0.075 G. In contrast, the RMSE of the combined estimator is 0.023 G. This indicates that the average improvement in mean square error (MSE) exceeds 10 dB.

\newpage
\section*{Conclusions}
In this study, we explored the integration of NV centers in bulk diamond with a Rb vapor cell, showcasing the benefits of both NV and Rb systems through comparisons of spatial resolution, field sensitivity, angular accuracy, and scalar accuracy. The developed system serves as an advanced platform for magnetic sensing, addressing both spatial and angular aspects. We investigated the enhanced estimation of vector fields, and the confocal system presents promising pathways for future gradiometric applications. By integrating high spatial resolution from the confocal setup, we have laid the groundwork for advanced spatial estimations.

Our focus was on estimating the vector field by leveraging the unique properties of Rb vapor and NV centers. Through simulations and theoretical analysis, we demonstrated that, in a shielded environment, substantial improvements in the accuracy of magnetic field magnitude estimation are achievable. Moreover, the precision of angular estimation relies heavily on the directional accuracy of NV measurements, a theme emphasized throughout this work.

For cases where the background field magnitude is much smaller compared to the target magnetic field, we showed that the improvement in estimation scales with the sensitivity ratio of the Rb system to the NV system, potentially leading to significant gains in measurement accuracy.

The performance of the integrated system was demonstrated experimentally, with a focus on illustrating an approach to system integration rather than achieving state-of-the-art sensitivity. We achieve a relative improvement of over 10 dB in the combined sensitivity compared to each sensor alone. We expect that in future experiments, by optimizing the system —enhancing both Rb and NV sensitivities through technical refinements like noise reduction, light collection improvements, and adjustments in microwave and laser power — we could notably boost the system's performance without altering the integrated setup's structure \cite{Budker2007, Taylor2008, Dang2010, Barry2020}.

Our experimental results aligned closely with simulations of the combined estimation, suggesting that such an integrated system holds substantial potential for practical applications. Moreover, the integration of two fundamentally distinct quantum platforms — solid-state NV centers and atomic Rb vapor — opens the door to a broad range of novel sensing modalities that leverage the complementary strengths of both systems.

\clearpage 
\bibliography{references} 

\begin{thebibliography}{10}
\providecommand{\url}[1]{\texttt{#1}}
\expandafter\ifx\csname urlstyle\endcsname\relax
  \providecommand{\doi}[1]{doi:\discretionary{}{}{}#1}\else
  \providecommand{\doi}{doi:\discretionary{}{}{}\begingroup \urlstyle{rm}\Url}\fi

\bibitem{Budker2007}
D.~Budker, M.~Romalis, Optical magnetometry. \emph{Nature Physics} \textbf{3}~(4), 227--234 (2007), \doi{10.1038/nphys566}.

\bibitem{Degen2017}
C.~L. Degen, F.~Reinhard, P.~Cappellaro, Quantum sensing. \emph{Reviews of Modern Physics} \textbf{89}~(3), 035002 (2017), \doi{10.1103/RevModPhys.89.035002}.

\bibitem{Hall2014_ReviewMagnetometry}
L.~T. Hall, J.~H. Cole, C.~D. Hill, L.~C.~L. Hollenberg, Sensing of Fluctuating Nanoscale Magnetic Fields Using Nitrogen-Vacancy Centers in Diamond. \emph{Phys. Rev. Lett.} \textbf{103}, 220802 (2009), \doi{10.1103/PhysRevLett.103.220802}.

\bibitem{Doherty2013}
M.~W. Doherty, \emph{et~al.}, The nitrogen-vacancy colour centre in diamond. \emph{Physics Reports} \textbf{528}~(1), 1--45 (2013), \doi{10.1016/j.physrep.2013.02.001}.

\bibitem{Rondin2014}
L.~Rondin, \emph{et~al.}, Magnetometry with nitrogen-vacancy defects in diamond. \emph{arXiv preprint arXiv:1311.5214}  (2014), \url{https://arxiv.org/abs/1311.5214}.

\bibitem{Taylor2008}
J.~M. Taylor, \emph{et~al.}, High-sensitivity diamond magnetometer with nanoscale resolution. \emph{Nature Physics} \textbf{4}~(10), 810--816 (2008), \doi{10.1038/nphys1075}, \url{https://arxiv.org/abs/0805.1367}.

\bibitem{Maze2008}
J.~R. Maze, \emph{et~al.}, Nanoscale magnetic sensing with an individual electronic spin in diamond. \emph{Nature} \textbf{455}~(7213), 644--647 (2008), \doi{10.1038/nature07279}.

\bibitem{Strner2021}
F.~M. Stürner, \emph{et~al.}, Integrated and Portable Magnetometer Based on Nitrogen‐Vacancy Ensembles in Diamond. \emph{Advanced Quantum Technologies} \textbf{4} (2021), \doi{10.1002/qute.202000111}.

\bibitem{Barry2024}
J.~F. Barry, \emph{et~al.}, Sensitive ac and dc magnetometry with nitrogen-vacancy-center ensembles in diamond. \emph{Physical Review Applied} \textbf{22}, 044069 (2024), \doi{10.1103/PhysRevApplied.22.044069}.

\bibitem{Wolf2015}
T.~Wolf, \emph{et~al.}, Subpicotesla Diamond Magnetometry. \emph{Physical Review X} \textbf{5}, 041001 (2015), \doi{10.1103/PhysRevX.5.041001}.

\bibitem{Barry2020}
J.~F. Barry, \emph{et~al.}, Sensitivity optimization for NV-diamond magnetometry. \emph{Reviews of Modern Physics} \textbf{92}~(1), 015004 (2020), \doi{10.1103/RevModPhys.92.015004}.

\bibitem{Grinolds2013_NVdiffraction}
M.~S. Grinolds, \emph{et~al.}, Nanoscale magnetic imaging of a single electron spin under ambient conditions. \emph{Nature Physics} \textbf{9}, 215--219 (2013), \doi{10.1038/nphys2543}.

\bibitem{Kominis2003}
I.~K. Kominis, T.~W. Kornack, J.~C. Allred, M.~V. Romalis, A subfemtotesla multichannel atomic magnetometer. \emph{Nature} \textbf{422}~(6932), 596--599 (2003), \doi{10.1038/nature01484}.

\bibitem{Seltzer2008}
S.~J. Seltzer, \emph{Developments in Alkali-Metal Atomic Magnetometry}, Ph.d. dissertation, Princeton University (2008), \url{https://ui.adsabs.harvard.edu/abs/2008PhDT.......309S/abstract}, aDS Bibcode: 2008PhDT.......309S.

\bibitem{Dang2010}
H.~B. Dang, A.~C. Maloof, M.~V. Romalis, Ultrahigh sensitivity magnetic field and magnetization measurements with an atomic magnetometer. \emph{Applied Physics Letters} \textbf{97}~(15), 151110 (2010), \doi{10.1063/1.3491215}, \url{https://pubs.aip.org/aip/apl/article/97/15/151110/122424/Ultrahigh-sensitivity-magnetic-field-and}.

\bibitem{Knappe2005_MicrofabricatedCells}
S.~Knappe, \emph{et~al.}, A novel technique for microfabricating alkali-atom vapor cells. \emph{Optics Letters} \textbf{30}~(18), 2351--2353 (2005), \doi{10.1364/OL.30.002351}, \url{https://opg.optica.org/ol/abstract.cfm?uri=ol-30-18-2351}.

\bibitem{Shah2007}
V.~Shah, S.~Knappe, P.~D.~D. Schwindt, J.~Kitching, Subpicotesla atomic magnetometry with a microfabricated vapor cell. \emph{Nature Photonics} \textbf{1}~(11), 649--652 (2007), \doi{10.1038/nphoton.2007.201}, \url{https://www.nature.com/articles/nphoton.2007.201}.

\bibitem{Sheng2013_RbVector}
B.~Patton, E.~Zhivun, D.~C. Hovde, D.~Budker, All‑Optical Vector Atomic Magnetometer. \emph{Physical Review Letters} \textbf{113}~(1), 013001 (2014), \doi{10.1103/PhysRevLett.113.013001}, \url{https://journals.aps.org/prl/abstract/10.1103/PhysRevLett.113.013001}.

\bibitem{Bell1961}
W.~E. Bell, A.~L. Bloom, Optical Detection of Magnetic Resonance in Alkali Metal Vapor. \emph{Physical Review} \textbf{107}~(6), 1559--1565 (1957), \doi{10.1103/PhysRev.107.1559}, \url{https://journals.aps.org/pr/abstract/10.1103/PhysRev.107.1559}.

\bibitem{Wolf2021_HybridSensing}
C.~Chia, D.~Huang, V.~Leong, J.~F. Kong, K.~E.~J. Goh, Hybrid quantum systems with artificial atoms in solid state. \emph{arXiv preprint arXiv:2404.05174}  (2024), \url{https://arxiv.org/abs/2404.05174}.

\bibitem{BarGill2021}
Y.~Ninio, \emph{et~al.}, High‑Sensitivity, High‑Resolution Detection of Reactive Oxygen Species Concentration Using NV Centers. \emph{ACS Photonics} \textbf{8}~(7), 1917--1921 (2021), \doi{10.1021/acsphotonics.1c00431}, \url{https://pubs.acs.org/doi/10.1021/acsphotonics.1c00431}.

\bibitem{Stern2023}
K.~Levi, A.~Giat, L.~Golan, E.~Talker, L.~Stern, Remote Chip-Scale Quantum Sensing of Magnetic Fields. \emph{Optica Quantum} \textbf{3}~(1), 84 (2024), \doi{10.1364/OPTICAQ.3.000084}, \url{https://opg.optica.org/opticaq/fulltext.cfm?uri=opticaq-3-1-84&id=567689}.

\bibitem{Pham2011}
L.~M. Pham, \emph{et~al.}, Magnetic field imaging with nitrogen-vacancy ensembles. \emph{New Journal of Physics} \textbf{13}~(4), 045021 (2011), \doi{10.1088/1367-2630/13/4/045021}, \url{https://arxiv.org/abs/1207.3339}.

\bibitem{Gruber1997}
A.~Gruber, \emph{et~al.}, Scanning Confocal Optical Microscopy and Magnetic Resonance on Single Defect Centers. \emph{Science} \textbf{276}~(5321), 2012--2014 (1997), \doi{10.1126/science.276.5321}, \url{https://www.science.org/doi/10.1126/science.276.5321.2012}.

\bibitem{Jelezko2006}
F.~Jelezko, J.~Wrachtrup, Single defect centres in diamond: A review. \emph{Physica Status Solidi A} \textbf{203}~(13), 3207--3225 (2006), \doi{10.1002/pssa.200671403}, \url{https://onlinelibrary.wiley.com/doi/10.1002/pssa.200671403}.

\bibitem{Scofield1994}
J.~H. Scofield, Frequency‑Domain Description of a Lock‑in Amplifier. \emph{American Journal of Physics} \textbf{62}~(2), 129--133 (1994), \doi{10.1119/1.17629}, \url{https://doi.org/10.1119/1.17629}.

\bibitem{Bevington2003}
P.~R. Bevington, D.~K. Robinson, \emph{Data Reduction and Error Analysis for the Physical Sciences} (McGraw‑Hill, New York, NY), 3rd ed. (2003), \url{https://cds.cern.ch/record/1305448}, includes bibliographical references and index.

\bibitem{Glenn2018}
D.~R. Glenn, \emph{et~al.}, High‑resolution magnetic resonance spectroscopy using a solid‑state spin sensor. \emph{Nature} \textbf{555}~(7696), 351--354 (2018), \doi{10.1038/nature25781}, \url{https://www.nature.com/articles/nature25781}.

\bibitem{methods}
Materials and methods are available as supplementary material.

\bibitem{Pham2013}
L.~M. Pham, \emph{Magnetic Field Sensing with Nitrogen-Vacancy Color Centers in Diamond}, Doctoral dissertation, Harvard University, Cambridge, MA (2013), \url{https://dash.harvard.edu/handle/1/11051173}, thesis (Ph.D.)—Harvard University, 2013.

\bibitem{Rb_Breit_Rabi}
B.~Wu, \emph{et~al.}, Experimental verification of the Breit–Rabi formula in the case of clock transition by using the spectroscopy method. \emph{Journal of the Optical Society of America B} \textbf{31}~(4), 742--747 (2014), \doi{10.1364/JOSAB.31.000742}, \url{https://opg.optica.org/josab/abstract.cfm?uri=josab-31-4-742}.

\end{thebibliography}
\bibliographystyle{sciencemag}

\section*{Acknowledgments}

\paragraph*{Funding:}
No external funding was received for this work.

\paragraph*{Author contributions:}
All authors contributed to the conception, development, and writing of the manuscript. Assistance from AI tools (specifically large language models) was used to help improve the clarity and phrasing of the text, under the authors' direction and supervision.

\paragraph*{Competing interests:}
There are no competing interests to declare.

\paragraph*{Data and materials availability:}
All data and materials used in this study are available from the corresponding author, I. Shalev (ittai.shalev@mail.huji.ac.il), upon reasonable request.


\subsection*{Supplementary materials}
Materials and Methods\\
Supplementary Text\\
Figs. S1 to S8\\
Equations S1 to S4\\
References \textit{(7-\arabic{enumiv})}\\ 


\begin{center}
    
\section*{Supplementary Materials for\\ \scititle}
I.~Shalev$^{\ast}$,
K.~Levi$^{}$,
R.~Malkinson$^{}$,
A.~Hen$^{}$,
L.~Stern$^{}$,
N.~Bar-Gill$^{}$

\small$^\ast$Corresponding author. Email: ittai.shalev@mail.huji.ac.il
\end{center}

\subsubsection*{This PDF file includes:}
Materials and Methods\\
Supplementary Text\\
Figs. S1 to S8\\
Equations S1 to S4\\
\renewcommand{\thefigure}{S\arabic{figure}}
\renewcommand{\thetable}{S\arabic{table}}
\renewcommand{\theequation}{S\arabic{equation}}
\renewcommand{\thepage}{S\arabic{page}}
\setcounter{figure}{0}
\setcounter{table}{0}
\setcounter{equation}{0}
\setcounter{page}{1} 

\newpage

\section*{Materials and Methods}

\subsection*{Rubidium and NV measurements and sensitivities}

\subsubsection*{NV center magnetic sensing and sensitivity}
The NV centers ODMR scan results in a spectrum that represents the projection of the magnetic field along each of the crystallographic orientations \( a \), \( b \), \( c \), and \( d \) \cite{Doherty2013, Rondin2014}.

The diamond utilized for the measurements is cut along the \([100]\) crystallographic direction. By knowing the crystallographic orientations \( a, b, c, d \), we can determine the expected frequency resonances of the various orientations on the ODMR scan, based on the specified requirements \cite{Maze2008, Barry2020}:

\begin{equation}
\begin{pmatrix}
{B}_a \\
{B}_b \\
{B}_c \\
{B}_d
\end{pmatrix}
=
\begin{pmatrix}
- {a}^T - \\
- {b}^T - \\
- {c}^T - \\
- {d}^T -
\end{pmatrix}
\begin{pmatrix}
{B}_x \\
{B}_y \\
{B}_z
\end{pmatrix}
\label{eq:NV_orientations_matrix}
\end{equation}
where the directions \( a, b, c, d \) are known relative to the laboratory coordinate system.

We note that the expression in Eq.~(\ref{eq:NV_orientations_matrix}) pertains to the measurement accuracy in relation to the diamond's orientations. To determine the measurement accuracy along the \( x \), \( y \), and \( z \) axes, we must apply a linear transformation of the uncertainty using the transition matrix \( W \), which converts the magnetic field measurements from the \( a, b, c, d \) orientations to the \( x, y, z \) coordinate system \cite{Pham2011}. It is also worth mentioning that we may choose to utilize only the three orientations with the highest sensitivity, characterized by high contrast and narrow bandwidth. Ultimately, this transformation yields the sensitivity in the \( x \), \( y \), and \( z \) directions:

\begin{equation}
\Delta  \mathbf {B}_{\text{NV}} = W \cdot \Delta \mathbf{B'}_{\text{NV}}
\label{eq:delta_B_NVi}
\end{equation}

\subsubsection*{Measured magnetic field sensitivities}
In the experiment of the marginal improvement distribution, we used a LIA chirp lasting 2.5 seconds, covering modulation frequencies from 300 to 1,500 kHz. For the ODMR measurements of the NVs, we performed 100 repeats with 15 averages, and for background field calculations, we used 150 averages. The scan included 60 frequencies within a 200 MHz range, with each single ODMR measurement per repeat and per average lasting 0.5 milliseconds. As described in previous chapters, it is possible to extract both the magnetic field measured by each system and the sensitivity of each system from these measurements \cite{Pham2013, Barry2020}.

The magnetic field we measured was generated by an applied current, with the induced field varying between 0 and 1.6 G.

In total, we conducted \( N = 50 \) repetitions for each sequence and each applied magnetic field to be measured. The estimated magnetic field is determined as the average of all repetitions, while the variance for each measurement is calculated using the empirical variance of the \( N \) measurements.

In the first step of the combined estimation algorithm, the background field \( \vec{B}_0 \) must be measured. Applying the algorithm outlined earlier, we measured \( \vec{B}_0 = (0.004, -0.7454, 0.6451) \, \text{G} \).

In the next step, we aim to determine the uncertainty associated with each measurement. To find the uncertainties for the NV system, we utilize Eqs.~\ref{eq:delta_B_NV} and~\ref{eq:delta_B_NVi}, which require knowledge of the error in the photoluminescence (PL) of the ODMR measurements—information obtained during the ODMR experiment—and the slope of the ODMR signal \cite{Pham2011}.

After calculating the slope of the ODMR and incorporating the PL error, we determined that the average uncertainty of the NV measurement was 260 mG.

\begin{figure}
    \centering
    \includegraphics[width=0.8\linewidth]{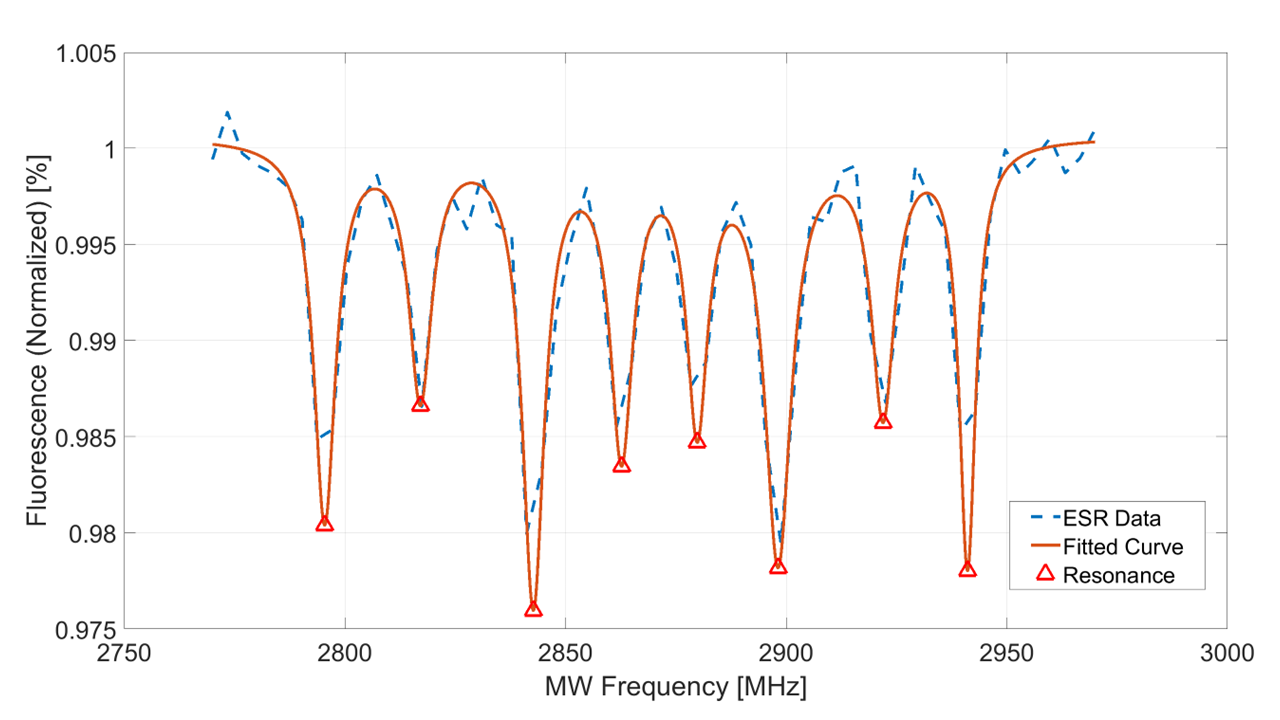}
    \caption{\textbf{ODMR experiment of the NV centers.} The ODMR experiment of measuring the NV system yielded an average slope of \( m_{\text{NV}} = 1.4 \cdot 10^{-3} \, \text{1/MHz} \), with an uncertainty in the photoluminescence (PL) given by \( \Delta PL = 0.6 \cdot 10^{-3} \). Consequently, the average uncertainty at the NV orientations is expressed as 
    \( \Delta B'_{\text{NV}} = \frac{\Delta PL}{\gamma_{\text{NV}} m_{\text{NV}}} = 150 \, \text{mG}, \)
    while the uncertainty in the XYZ reference system is given by 
    \( \Delta B_{(\text{NV}, \hat{e}_i)} = W_{\hat{e}_i} \cdot \Delta \vec{B}'_{\text{NV}} = 260 \, \text{mG}. \)}
    \label{fig:ODMR}
\end{figure}

Similarly, we aim to understand the uncertainty associated with the Rb measurement. To achieve this, we utilize Eq.~\ref{eq:delta_B_Rb}, which requires us to measure the standard deviation of the LIA at the maximum of the Y measurement \cite{Seltzer2008, Budker2007}.

The noise calculation of the Y signal is performed through linear fitting, measuring the root mean square error (RMSE) of the LIA signal, with the slope derived from the linear fitting process. Using Eq.~(13), we determined that the uncertainty in measuring the magnetic field with the Rb system is \( 790 \, \mu \text{G} \). This linear approximation neglects the small but non-negligible higher-order Zeeman shifts, which can introduce frequency offsets on the order of tens of Hertz. Incorporating the full Breit–Rabi solution \cite{Rb_Breit_Rabi} into the frequency–field calibration captures these nonlinearities and can further enhance measurement accuracy.

\begin{figure}
    \centering
    \includegraphics[width=0.8\linewidth]{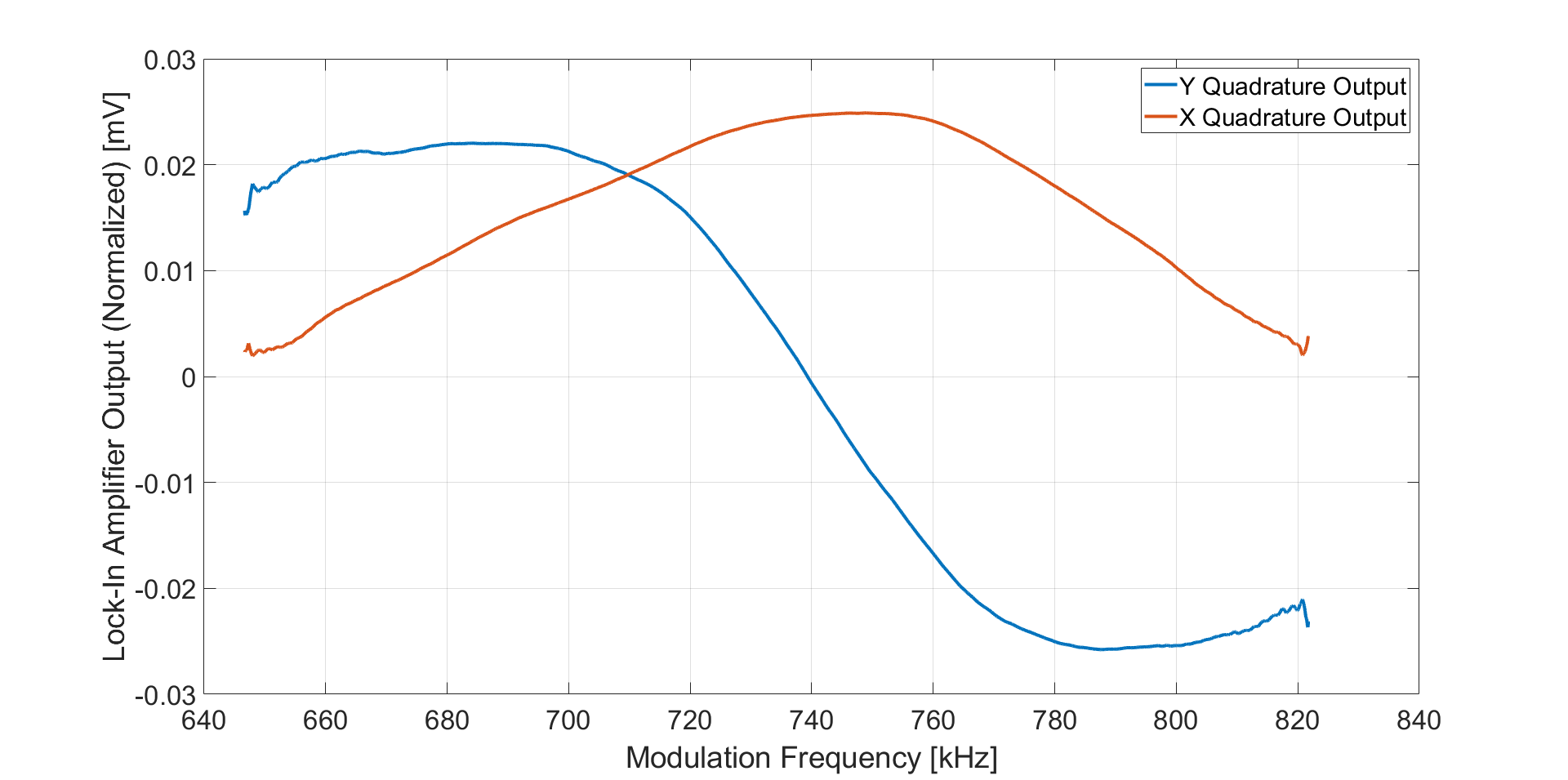}
    \caption{\textbf{LIA measurement of the Rb system}. The average slope of the Y output is \( m_{\text{Rb}} =  10^{-6} \, \text{V/kHz} \), where the uncertainty of the Y signal is \( \Delta Y = 5.5 \times 10^{-6} \, \text{V} \). Consequently, the uncertainty of the measurement is given by
    \( \Delta B_{\text{Rb}} = \frac{\Delta f}{\gamma_{\text{Rb}}} = \frac{\Delta Y}{\gamma_{\text{Rb}} m_{\text{Rb}}} = 790 \, \mu\text{G}. \)}
    \label{fig:LIA_signal}
\end{figure}

\subsection*{Setup}

The core component of the combined setup is the integrated sensor, which enables magnetic field sensing at a defined spatial location and time. This sensor consists of a rubidium (Rb) vapor cell positioned adjacent to a bulk diamond, with polyimide spacers maintaining a fixed separation between the two components.

The Rb cells are fabricated by anodically bonding Pyrex glass layers to both sides of a silicon substrate, which includes a through-hole to define the internal vapor volume \cite{Seltzer2008}. This volume is filled with $\mathrm{^{87}Rb}$ vapor and a buffer gas mixture of nitrogen and argon in a 2:1 ratio, pressurized to 40 Torr. The buffer gas serves to prolong ground-state lifetimes and suppress fluorescence \cite{Seltzer2008}. The complete cell dimensions are 5 mm × 5 mm × 3 mm, with an internal cylindrical vapor chamber measuring 1.5 mm in diameter and 2.4 mm in length.

To increase the rubidium vapor density, the cell must be heated. This is achieved by irradiating the silicon substrate with a 450 nm blue laser, which induces a localized temperature increase. Our previous work employing a comparable micromachined vapor cell with remote atomic density control achieved a sensitivity of ~ \( \text{1pT}/\sqrt{\text{Hz}} \) in an unshielded environment at a stand-off distance of 10 m \cite{Stern2023}.

Two 0.6 mm-thick polyimide spacers are attached to one of the transparent surfaces of the Rb cell. A bulk diamond containing nitrogen-vacancy (NV) centers is mounted beneath these spacers. The spacers provide thermal insulation, confining the heat generated by the laser to the Rb cell, thereby facilitating efficient vapor generation without excessive heat transfer to the diamond.

The diamond is a high-purity bulk crystal with a $\mathrm{^{15}N}$ concentration of 25 parts per million (ppm), distributed within a depth of approximately 30~$\mu$m from the surface. The crystal dimensions are 5 mm × 5 mm × 0.5 mm. When used with a confocal microscope and without lock-in detection, the NV-based system achieves a sensitivity of approximately \( \mu\text{T}/\sqrt{\text{Hz}} \) \cite{Pham2013}.

The diamond is mounted on a coplanar microwave antenna. This antenna is fabricated on a 20 mm × 20 mm, 100~$\mu$m-thick glass substrate with printed copper traces 40~$\mu$m wide and spaced 40~$\mu$m apart. It enables the application of microwave fields to drive optically detected magnetic resonance (ODMR) in the NV centers \cite{Pham2013, Barry2020}.

The antenna is connected via conductive pads to a printed circuit board (PCB) equipped with BNC connectors, which deliver microwave signals from a waveform generator. The PCB is mounted on a dual-stage positioning system: one stage with 1~$\mu$m resolution and a second with sub-micron precision. This setup allows confocal imaging and precise alignment of the sensor with respect to the optical axis.

The optical system enables independent excitation and detection of both the NV and Rb sensors using multiple laser sources. A high-numerical-aperture oil-immersion objective lens (NA = 1.40, 60X), placed beneath the antenna, focuses a 532~nm green laser for NV excitation and a 795~nm red laser for Rb excitation. The green beam is pulsed using an acousto-optic modulator (AOM) and directed through the first port of a 780~nm dichroic mirror (DM), which transmits wavelengths below 780~nm and reflects longer wavelengths. This optical configuration separates the NV and Rb excitation paths, allowing simultaneous operation without cross-interference \cite{Pham2013, Seltzer2008}.

The 795~nm red laser, modulated by a signal generator, is circularly polarized, expanded, and focused using a lens assembly to uniformly illuminate the Rb cell. A neutral density (ND) filter reduces the beam intensity to the order of tens of microwatts. 
Red fluorescence (640–800~nm) emitted by the NV centers is collected through the objective and separated from the excitation path by a second dichroic mirror, which transmits red and reflects green wavelengths. The fluorescence is then directed to a single-photon counting module (SPCM)\cite{Pham2013}.

\begin{figure}
    \centering
    \includegraphics[width=0.8\linewidth]{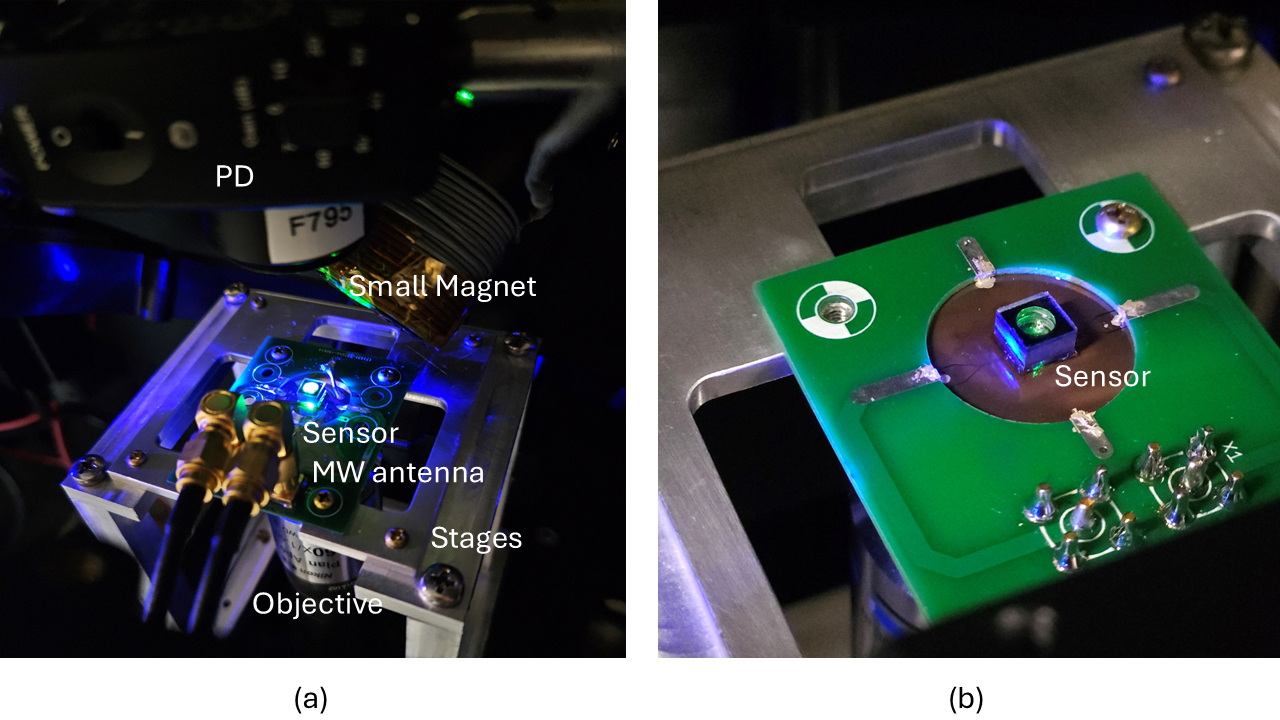}
    \caption{\textbf{Overview of the combined sensor.} \textbf{(a)} Configuration of the sensing module during measurement. \textbf{(b)} The integrated sensor—comprising the Rb cell and NV diamond mounted on a microwave antenna—is connected to a PCB and positioned above the objective. The entire sensor assembly is mounted on precision stages for accurate spatial alignment.}
    \label{fig:real_sensor}
\end{figure}

\section*{Supplementary Text}

\subsection*{Angular Accuracy}
\begin{figure}[tbh]
    \centering
    \includegraphics[width = 0.6 \linewidth]{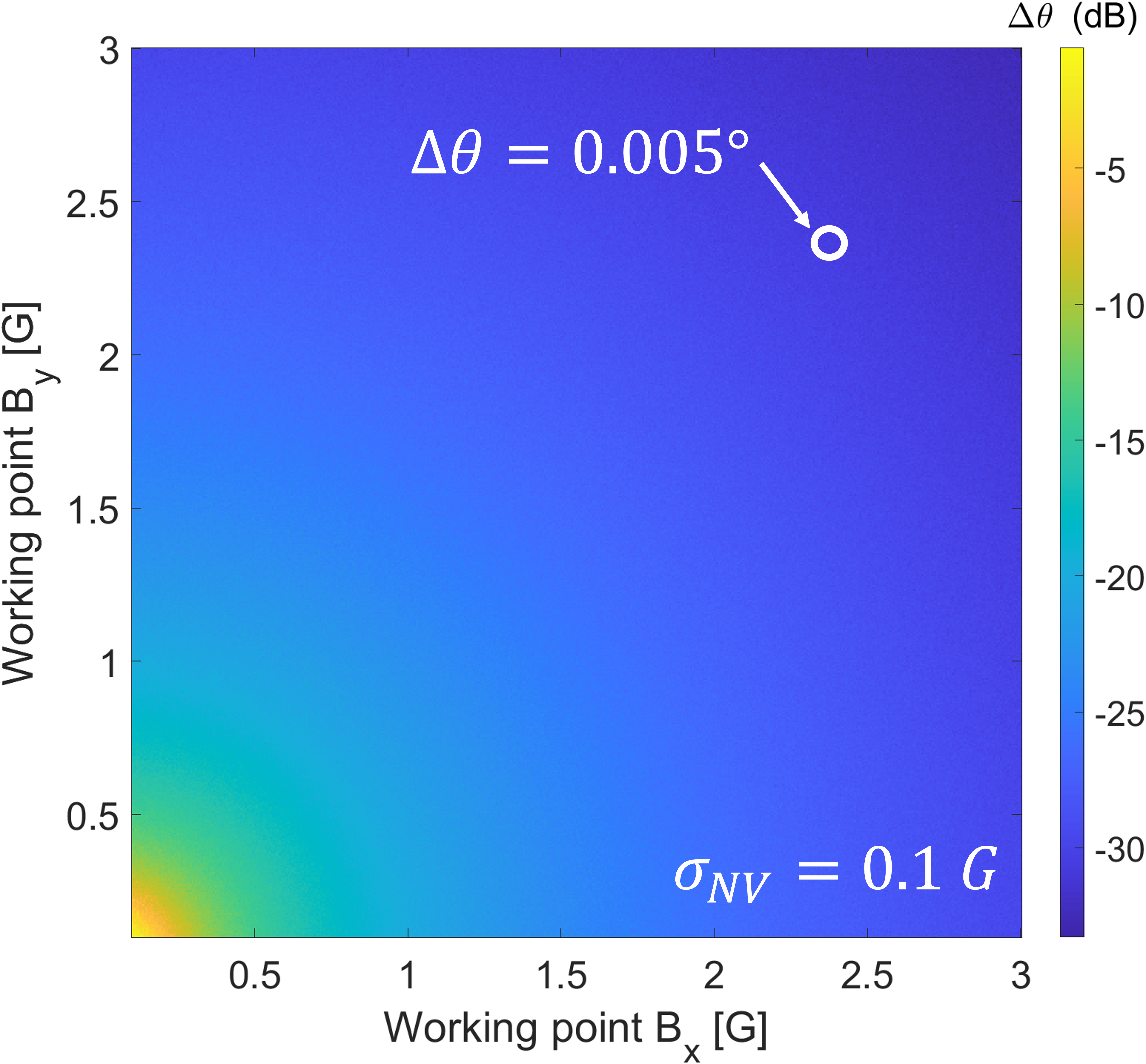}
    \caption{\textbf{Simulation of the uncertainty in the angle of the NV measurement in dB scale, as a function of the magnetic field measured at the x-y plane, given a measurement uncertainty of 0.1 G.} The larger the field, the smaller the angular error, meaning that by using a higher magnetic field magnitude working point it is possible to decrease the angular uncertainty.}
    \label{fig:angular_error}
\end{figure}

Let us consider the scenario where we aim to measure a magnetic field oriented in the $\theta$) direction (or $\mathbf{\Theta} = (\theta, \varphi)$ for the general case), characterized by a magnitude \( {B} \). Analyzing  the NV system alone, For small magnetic fields, for which magnitude error of the measurement is comparable to the magnitude of the field being measured, the uncertainty in the angle will be maximal. Conversely, as the magnitude of the magnetic field increases relative to the error, the uncertainty in the angle decreases. In contrast, the uncertainty in the magnitude of the field remains constant. The relationship of the NV cartesian measurement uncertainty and the direction uncertainty is given by the following equations:

\begin{equation}
\theta_B + \Delta \theta_B = \arctan \left( \frac{\sqrt{(B_x + \Delta B_x)^2 + (B_y + \Delta B_y)^2}}{B_z + \Delta B_z} \right)
\end{equation}

\begin{equation}
\varphi_B + \Delta \varphi_B = \arctan \left( \frac{B_y + \Delta B_y}{B_x + \Delta B_x} \right)
\end{equation}
where \(\theta_B\) and \(\varphi_B\) represent the deterministic true directions of the measured magnetic field, and \(\Delta \theta_B\), \(\Delta \varphi_B\), \(\Delta B_i\) correspond to the uncertainties in the angles and Cartesian components of the magnetic field, which are treated as random variables. When the uncertainty in the NV measurement \(\Delta B_i\) is significantly smaller than the true magnitude of the magnetic field, it follows that the uncertainties in the magnetic field directions, \(\Delta \theta_B\) and \(\Delta \varphi_B\), will also be reduced. This results in an enhanced accuracy for the determination of the magnetic field's direction, as illustrated in Fig. \ref{fig:angular_error}.

\subsection*{Shielded Environment Measurement Simulation}
A simulation that estimates the magnetic field using both the combined system and the NV system independently is presented. In this simulation, we defined the uncertainty of the NV system to be 1,000 times greater than that of the Rb system. The magnetic field was estimated within the range of -1.5 G to 1.5 G, with the deterministic field constrained to the x-y plane, while noise was present in three dimensions. Each field value was tested through 50 repetitions to ensure sufficient statistical data. Subsequently, we calculated the improvement $G$ of the combined system's measurements compared to those obtained using the NV system alone, assessing both the magnitude and direction of the magnetic field. The improvement is quantified by the ratio of the mean square error (MSE), and mean absolute error (MAE), between the combined estimator and the NV-only measurement, expressed on a logarithmic scale (dB), for each simulated magnetic field.

\begin{figure}[tbh]
    \centering
    \includegraphics[width = 1 \linewidth]{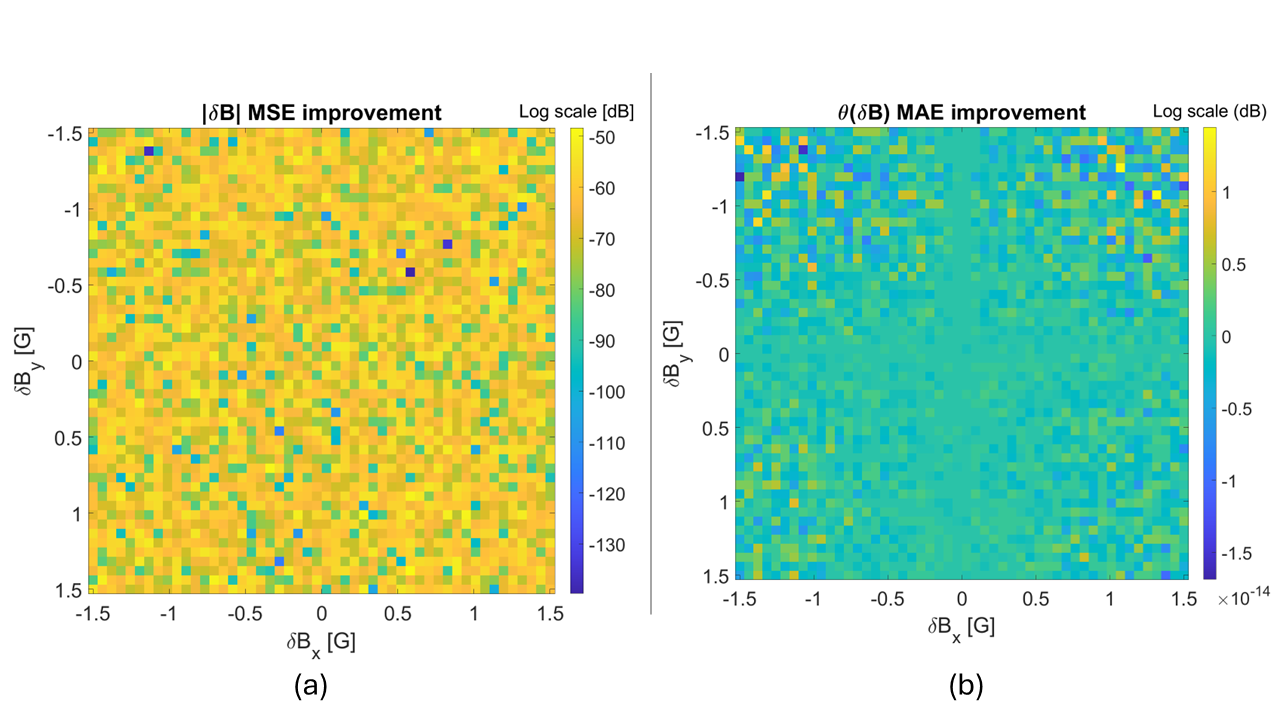}
    \caption{\textbf{The improvement in estimating both the magnitude (plot \textbf{(a)}) and the direction (plot \textbf{(b)}) of the magnetic field using the combined estimator, as compared to the NV system alone, with no backround field applied.} Notably, in terms of mean square error (MSE), the enhancement in the estimation of the magnetic field magnitude is 60 dB, indicating that the standard deviation of the combined system is improved by approximately 30 dB relative to the NV measurement alone. Conversely, there is no observed improvement in the estimation of the magnetic field direction.}
    \label{fig:simulation_no_B0}
\end{figure}

Under ideal conditions—where the background field is significantly smaller than the target field, or the system is calibrated in a shielded environment to minimize the background field's influence on the NV measurements—the accuracy of the magnetic field measurement can be greatly enhanced as shown in Fig. \ref{fig:simulation_no_B0}. The error in the combined estimator’s magnetic field magnitude will improve proportionally to the difference in the orders of magnitude between the NV and Rb measurement’s uncertainties. For instance, if the NV sensor has an accuracy of microteslas and the Rb sensor achieves nanotesla accuracy, the combined estimator will yield a 1,000-fold improvement in magnetic field precision compared to using the NV measurement alone.

Regarding angular accuracy, the combined measurement enhances only the Rb system’s performance. This is because it not only benefits from the same improvement in magnetic field magnitude accuracy but also provides an estimation of the field’s angle.

\subsection*{Enhanced Measurement At The Presence Of Background Magnetic Field}

\begin{figure}
    \centering
    \includegraphics[width=0.9\linewidth]{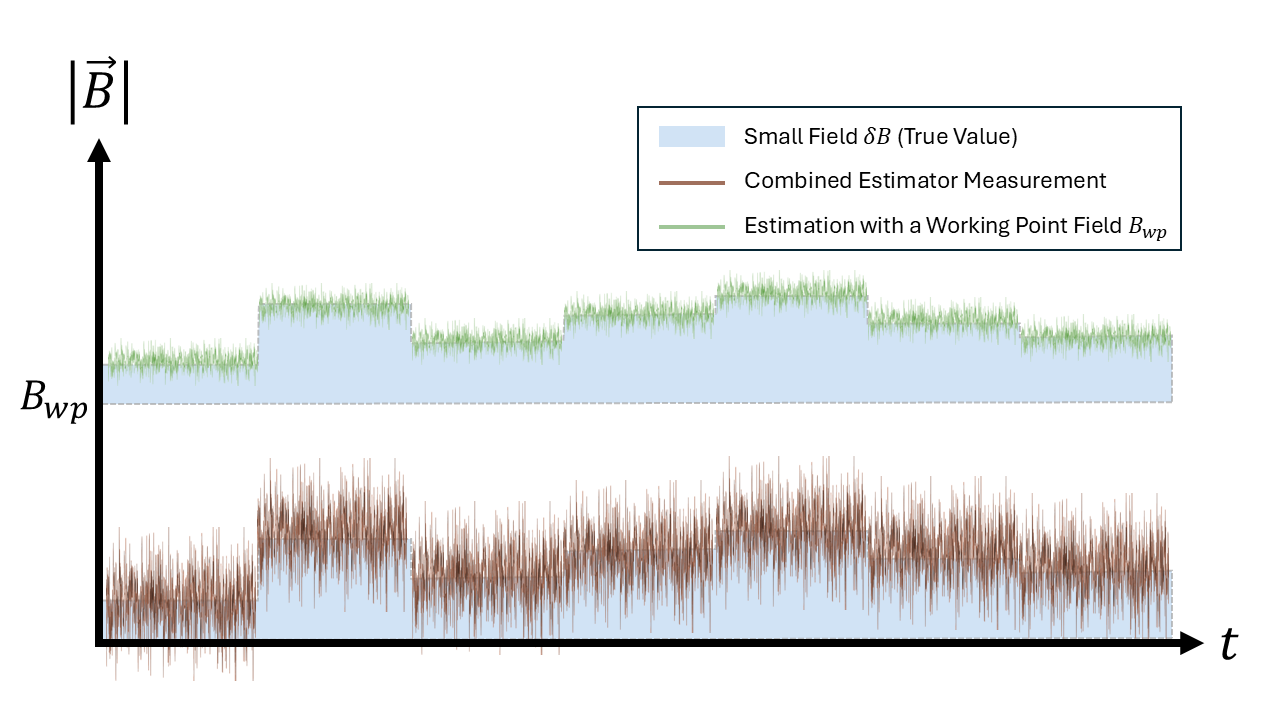}
    \caption{\textbf{Measurement of the magnetic field $\delta \vec{B}$ using the combined estimator when the background field $\vec{B}_0$ is significant}. The red signal indicates the measurements from the integrated system and their associated uncertainty without the application of the working point bias field $\vec{B}_{\text{wp}}$. In contrast, the green signal shows the combined system's measurements and their reduced uncertainty when the $\vec{B}_{\text{wp}}$ field is activated.}
    \label{fig:B_wp}
\end{figure}

As been shown, the presence of a significant background field reduces the performance of the combined estimator. However, to mitigate this issue, a controlled and stable external field $\vec {B}{_{wp}}$ can be applied in such a way that it is turned on during the Rb measurement, and during the NV measurement it is turned off during the reference measurement. This technique shifts the working point of the field estimation into a regime (according to Fig. (\ref{fig:simulation_with_B0})) where the performance of the combined estimator improves, as illustrated in Fig. \ref{fig:B_wp}.

\subsection*{Scalar Correction VS. Vector Correction}
\begin{figure}
    \centering
    \includegraphics[width=1\linewidth]{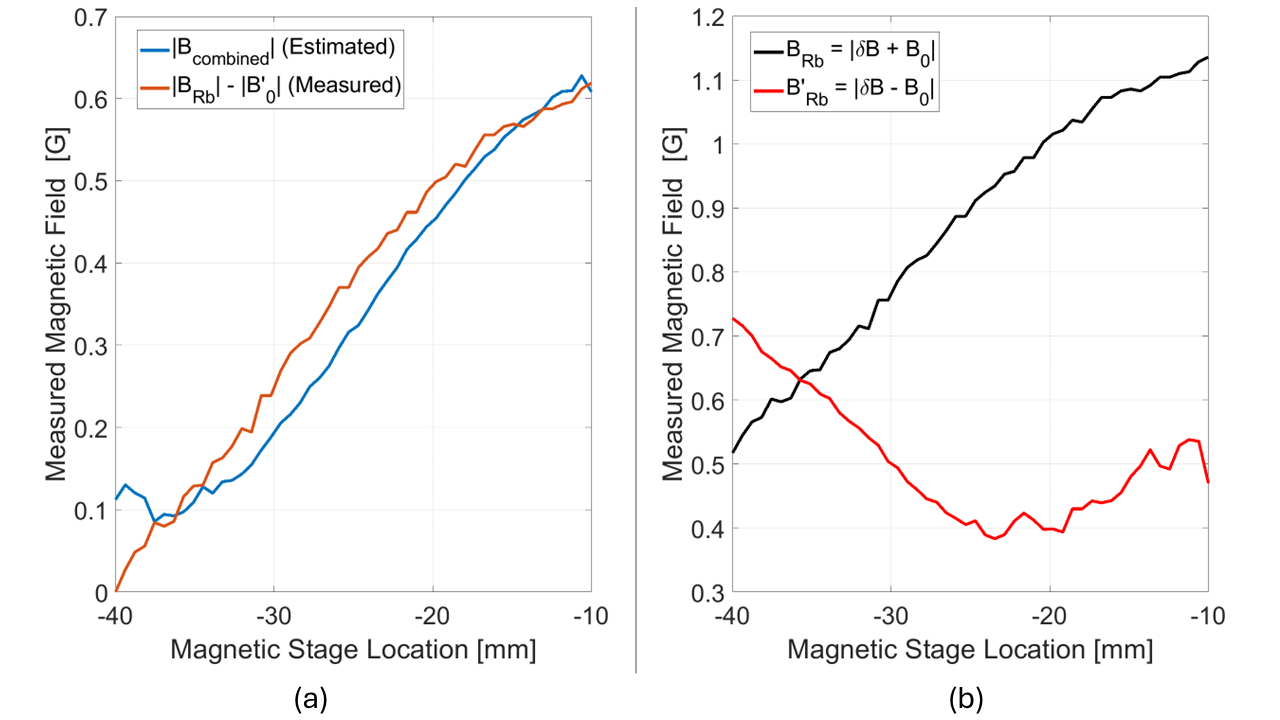 }
     \caption{\textbf{Comparison between hybrid estimation and a non-hybrid estimation method in the presence of a background field}. (a) Comparison between the combined estimator and the Rb measurement, in the case where we naively subtracted the scalar value of the background field from the rubidium measurement. It can be seen that the two curves have a different shape, due to the distortion that the background field introduces into the Rb measurement. (b) Measuring the magnetic field using the Rb system. In black, is the real Rb measurement performed during the experiment, while in red is a simulation of the Rb measurement if the background field $\vec{B}{_0}$ was in the opposite direction from the direction we measured. In both cases, the combined estimator would have given the same result (because an inherent part of it is the vector evaluation of $\vec{B}{_0}$), and in contrast, if we were to refer to the Rb measurement naively, it can be seen that for a different $\vec{B}{_0}$ we would have obtained completely different curves, even though the set-up of the magnet is the same in both cases.}
    \label{fig:distortion}
\end{figure}

It is important to notice that the estimation of the background field is critical for an accurate estimation. The Rb measurement is a scalar measurement that contains the background field, meaning that it mixes the background field with the actual desired field of the source. If the background field is not at the exact same direction of the magnetic field of the desired source, the Rb measurement curve will have an inherent distortion, as can be seen in Fig. \ref{fig:distortion} (a), and it is impossible to determine how to seperate the background field and the desired small field of interest.

In order to illustrate the distortion created by the background noise in a more significant way, we performed a simulation in which the background field is in the opposite direction to that measured in the experimental system, and we added it vectorially to the measured field as can be seen in Fig. \ref{fig:distortion} (b). In this way, it is possible to simulate how the rubidium measurement curve will be distorted rapidly as the direction of the background field changes.


\end{document}